\documentclass[twocolumn]{aastex}
\usepackage[english]{babel}
\usepackage{amsmath}
\usepackage{graphicx}
\usepackage{wasysym}
\usepackage{soul}
\usepackage{parskip}
\usepackage{wrapfig}
\usepackage{url}
\usepackage{float}
\newcommand{\mhz}{$\mathrm{\mu Hz}$}

\begin{document}
\pagestyle{myheadings}
\title{Dealing With $\delta$-Scuti Variables: Transit Light Curve Analysis of Planets Orbiting Rapidly-Rotating, Seismically Active A/F Stars}

\author{John P. Ahlers}
\affil{\emph{Physics Department, University of Idaho, Moscow, ID 83844}}
\affil{\emph{Exoplanets and Stellar Astrophysics Laboratory, Code 667, NASA Goddard Space Flight Center, Greenbelt, MD 20771, USA}}

\author{Jason W. Barnes}
\affil{\emph{Physics Department, University of Idaho, Moscow, ID 83844}}

\author{Samuel A. Myers} 
\affil{\emph{Physics Department, University of Idaho, Moscow, ID 83844}}

\begin{abstract}
We measure the bulk system parameters of the seismically active, rapidly-rotating $\delta$-Scuti KOI-976 and constrain the orbit geometry of its transiting binary companion using a combined approach of asteroseismology and gravity-darkening light curve analysis. KOI-976 is a $1.62\pm0.2~\mathrm{M_\odot}$ star with a measured $v\sin(i)$ of $120\pm2$ km/s and seismically-induced variable signal that varies by $\sim$ 0.6\% of the star's total photometric brightness. We take advantage of the star's oblate shape and seismic activity to perform three measurements of its obliquity angle relative to the plane of the sky. We first apply rotational splitting theory to the star's variable signal observed in short-cadence \emph{Kepler} photometry to constrain KOI-976's obliquity angle, and then subtract off variability from that dataset using the linear algorithm for significance reduction software {\tt LASR}. We perform gravity-darkened fits to \emph{Kepler} variability-subtracted short-cadence photometry and to \emph{Kepler's} phase-folded long-cadence photometry to obtain two more measurements of the star's obliquity. We find that the binary system transits in a grazing configuration with measured obliquity values of $36^\circ\pm17^\circ$, $46^\circ\pm16^\circ$, and $43^\circ\pm20^\circ$ respectively for the three measurements. We perform these analyses as a way to demonstrate overcoming the challenges high-mass stars can present to transit light curve fitting and to prepare for the large number of exoplanets \emph{TESS} will discover orbiting A/F stars.
\end{abstract}


\section{Introduction} \label{sec:intro}
High-mass stars of $\sim1.3M_\odot$ present unique challenges to transit light curve analysis that stem from their fundamental structure. In general, stars less massive than the Sun have larger convective zones and smaller radiative zones, and stars more massive than the Sun have larger radiative zones and smaller convective zones \citep{toomre1976stellar}. However, at masses higher than $\sim1.3M_\odot$ the star's carbon-nitrogen-oxygen cycle of nuclear fusion produces an extremely high core temperature, causing the core itself to become convective. This convective core resides inside a radiative region that extends to the star's surface. The result is that these stars are effectively inside-out from their low-mass counterparts, with convective interiors and radiative exteriors. \citet{2012ApJ...757...18A} identifies this inversion at a stellar surface temperature of $\sim6200\mathrm{K}$. Throughout this work, we use $M_\star\approx1.3M_\odot$ and $T_\mathrm{eff}\approx6200$ as approximate cutoffs for designating a star ``high-mass'' or ``low-mass''.

As a consequence of their structure, high-mass stars have weak external magnetic fields. Solar dynamo theory states that stellar magnetic fields are caused by the convection zone of the star \citep{charbonneau2014solar}, so the magnetic field in high-mass stars should be mostly internal near the star's convective core. Recent observations using NASA's \emph{Kepler} telescope corroborate this theory \citep[e.g.,][]{bagnulo2006searching,boehm2015discovery}. 

Without a strong external magnetic field, no stellar magnetic braking takes place, allowing high-mass stars to maintain their primordial rotation rates throughout their lifetimes \citep{mestel1968magnetic}. All stars start off spinning quickly during their formation as protostellar material collapses inward \citep{1994sipp}; however, low-mass stars' external magnetic fields cause them to slowly decrease their rate of rotation over time \citep{meibom2009stellar}. This effect, called magnetic braking, occurs when the star's magnetic field transfers angular momentum into an escaping stellar wind. These outflows are stirred by the star's magnetic field, which transfers angular momentum from the star to the outflow, slowing the star's rotation rate. However, angular momentum transfer does not occur between a high-mass star and its outflow, allowing them to stay spinning rapidly throughout their main sequence.

High mass stars' rotation rates often hover near their rotational break-up speed, with equatorial rotational velocities reaching hundreds of kilometers per second \citep[e.g.,][]{royer2002rotational, jackson2004models, huang2006stellar}. Their high rotation rates distort their shapes into oblate spheroids that bulge outward at the equator. For example, the well-known rapid rotator Altair spins at 72\% of its breakup speed and has an oblateness factor of 0.177, meaning Altair's polar radius is only 82.3\% as large as its equatorial radius \citep{2007Sci...317..342M}.

The smaller effective surface gravity near a rapid rotator's equator results in a lower stellar effective temperature. The von Zeipel theorem \citep{von1924radiative} or ``gravity-darkening law" relates surface gravity with effective temperature:

\begin{equation}
T_\mathrm{eff}=T_\mathrm{pole}\left(\frac{g_\mathrm{eff}}{g_\mathrm{pole}}\right)^\beta
\label{eq:GRAVDARKgravdark}
\end{equation}

where $\beta$ is the so-called gravity-darkening exponent. The star's rapid rotation displaces its hydrostatic equilibrium and induces a pole-to-equator temperature gradient across the stellar surface, resulting in poles that can be several thousand Kelvin hotter than the star's equator. As an example, Altair's stellar effective temperature varies from $\sim8500$K to $6500$K between its hot poles and cool equator \citep{kervella2005gravitational}.

Additionally, A/F stars can display photometric variability that can obfuscate transit light curve analysis. These stars pulsate with changing light amplitudes that can range from 0.003 to 0.8 magnitudes \citep{breger2000delta}, enough to drastically alter or even hide transit events. The most commonly observed variables in the A/F spectral classes are $\delta$-Scuti and their cousins $\gamma$-Doradus and dwarf Cepheid variables. NASA's \emph{Kepler} telescope observed over 1400 $\delta$-Scuti stars out of the $\sim$150,000 stars in \emph{Kepler's} field of view \citep{balona2012search}. 

Previous works have shown that stellar variability seen in $\delta$-Scuti can reveal useful information about a system. \citet{goupil2000rotational} shows that $\delta$-Scuti variables typically oscillate with only a few dominant frequencies, and that the separation between modes can reveal the star's rotation rate and its obliquity relative to the plane of the sky. \citet{zwintz2014refining}  applied second-order perturbation theory to the intermediate Delta Scuti star HD 144277. \citet{herrero2011wasp} and others \citep[e.g.,][]{smith2011thermal,de2013grouse,von2014pulsation} have previously combined asteroseismology and exoplanet analysis to characterize WASP-33, a $\delta$ Scuti hosting a transiting hot Jupiter. 

Multiple techniques exist for resolving a variable stellar signal in photometry. Previous works on $\delta$-Scuti modelled photometric variability using the iterative fitting process of fitting more and more sinusoids to the time series using linear regression, known as prewhitening \citep[e.g.,][]{machado2008ccd,breger2011rotational,breger2016nonradial}. In this work we apply a similar fitting routine, the Linear Algorithm for Significance Reduction \citep{ahlers2018lasr}, which subtracts stellar oscillations one at a time from photometry by using each oscillation's statistical significance in frequency space as a goodness-of-fit parameter.  

The Transiting Exoplanet Survey Satellite (TESS) expects to discover $\sim$ 2000 exoplanets orbiting A/F stars \citep{barclay2018revised}. Most of these exoplanets will likely orbit rapid rotators, and proper modelling of their transit light curves will need to include the gravity-darkening phenomenon \citep{barnes2009transit,ahlers2016gravity}. As many as $\sim$200 may also transit $\delta$-Scuti and their cousins \citep{balona2012search}. Therefore, proper handling of rapid stellar rotation and stellar seismic activity will enable the analysis of a large fraction of exoplanets discovered by \emph{TESS}.

This work demonstrates how to handle both rapid stellar rotation and stellar variability when extracting information from a transit light curve. We show that these phenomena are not merely complications to a model or noise to be subtracted from a signal, but that they can provide useful insight to the bulk properties of a system through the example system Kepler Object of Interest (KOI) 976. In \S2 we detail our asteroseismic approach and demonstrate the advantages of working with short-cadence photometry. In \S3 we show the results of our work. In \S4 we discuss expected results when performing these analyses on photometry from NASA's \emph{TESS} mission. 

\section{Methods}

\subsection{Observations}
In this work, we choose the $\delta$-Scuti Kepler Object of Interest (KOI) 976 -- KIC 3441784 -- as our target system for analysis. The Exoplanet Follow-Up Observing Program (ExoFOP) lists this system as an eclipsing binary consisting of an F0 star and an M-dwarf companion. The F0 star (hereafter called the primary star) is rotating rapidly with an ExoFOP-reported $v\sin(i)$ of $120~\mathrm{km/s}$ and seismically-induced amplitude variations of $\sim$3.0mmag. Addionally, ExoFOP identifies the primary star as a High-Amplitude $\delta$-Scuti variable (HAD); however, HADs typically have low $v\sin(i)$ and amplitude variations larger than 0.3mag \citep{pigulski2005high}. Thus we classify KOI-976 as a $\delta$-Scuti but not an HAD. \citet{baranec2016robo} identifies a binary companion (hereafter called the transiting star) that transits the primary star once every 52.6 days. We list all previously reported values of KOI-976 in Table \ref{table:parameters}.

\renewcommand{\arraystretch}{1.2}
\begin{table}[h] 
\centering
\begin{tabular}{lr}
\hline
\hline
{\bf Parameter} & {\bf Value} \\ \hline
$M_\star~(M_\odot)$ & $1.62\pm0.2$ \\
$T_\mathrm{eff}$~(K) & $7240\pm200$ \\
Kepler mag & $9.729$ \\
$v\sin(i)~(\mathrm{km/s})$  & $120\pm2$ \\
Transit Period (days) & $52.56902\pm5\times10^{-5}$ \\
Transit Depth (mmag) & $29.61\pm0.19$ \\

\end{tabular}
\caption{Previously reported values of KOI-976. $M_\star$, $T_\mathrm{eff}$, and $v\sin(i)$ were measured with the Trans-Atlantic Exoplanet Survey (TReS) telescope and are listed on the Exoplanet Follow-up Observing Program (ExoFOP).}
\label{table:parameters}
\end{table}

KOI-976 serves as an ideal candidate for this analysis for several reasons. The primary star is a poster-child for $\delta$-Scuti with a dominant oscillation period of $\sim1.1$ hours and total changes in the star's photometric brightness of $\sim$0.6\%. Like most $\delta$-Scuti, KOI-976's seismic activiy is dominated by only a few low-degree oscillations. ExoFOP lists the primary star's \emph{Kepler} magnitude as 9.7, providing an excellent signal-to-noise in both \emph{Kepler's} long-cadence and short-cadence datasets. Additionally, the transiting companion provides a transit depth of 30 mmag, making the transit event easy to resolve out of the primary star's variable signal.

In this analysis, we make use of both \emph{Kepler's} long-cadence and short-cadence photometric datasets of KOI-976. The 30-minute long-cadence photometry of KOI-976 spans NASA's primary \emph{Kepler} mission with Q0-Q17 observations of 27 transit events from May 2009 to April 2013. We include the entire Q0-Q17 long-cadence dataset in our analysis. \emph{Kepler's} 1-minute short-cadence photometry of KOI-976 includes Q8 and Q9 observations in early 2011. However, these datasets have a large observation gap between the two quarters, and since only the Q9 short-cadence dataset contains a transit, we do not include the Q8 dataset in this work. 

\subsection{Binary Time-Delay}\label{sec:timedelay}
\citet{balona2014binary} identified a previously-unknown binary companion in the KOI-976 system with a 208 day orbit period. They found that this third star in the system contains sufficient angular momentum to significantly shift the location of the system barycenter. They discovered this third star using the time-delay method of binary star detection, in which the $\delta$-Scuti's variable photometric signal undergoes phase shifts caused by a speed-of-light delay as is orbits about barycenter.

\citet{balona2014binary} measured that the primary star orbits about barycenter in an eccentric orbit once every 208 days with a semimajor axis of $0.27$ au, corresponding to a speed-of-light delay that changes by $\sim$200 seconds throughout the orbit. We account for this effect by adjusting both the long-cadence and short-cadence \emph{Kepler} datasets according to their time-delay results. We add/subtract time to each time bin in the datasets according to the primary star's location relative to the system barycenter. See \citet{balona2014binary} for the details of our time-delay adjustment.

\subsection{Asteroseismology} \label{sec:astero}
Our asteroseismic analysis of KOI-976 has two main goals: to constrain the primary star's rotation rate and obliquity angle by appling rotational splitting to its out-of-transit variable signal, and to prepare the photometric dataset for transit light curve analysis by subtracting off stellar variability. The highest-frequency oscillation that can be resolved in \emph{Kepler's} 30-minute long-cadence data is defined by its Nyquist rate of $277.8$~\mhz~\citep{sampford1962introduction}. However, we find in short-cadence that KOI-976 possesses significant frequencies up to $\sim800$~\mhz, so in our asteroseismic analysis we exclusively use \emph{Kepler's} 1-minute Q9 short-cadence photometry, which has a cutoff Nyquist rate of $8333.3$~\mhz.

\subsubsection{Rotational Splitting}
$\delta$-Scuti typically oscillate at a few dominant frequencies. These frequencies derive from variations in stellar shape that obey the physics of spherical harmonics,
\begin{equation}
Y^m_l(\theta,\phi) = \sqrt{\frac{(2l+1)(l-m)!}{4\pi(l+m)!}}P^m_l(\cos(\theta))e^{im\phi}
\end{equation}

where $Y$ describes the oscillation across the star's surface, defined by the azimuthal and polar angles $(\phi,\theta)$, degree $l$, its order $m$, and corresponding Legendre polynomial $P^m_l(\cos(\theta))$. For a given degree $l$, there are $2l+1$ oscillation modes, one for each $-l\leq m\leq l$ order. Together these grouped oscillations are known as a multiplet. Typically, $\delta$-Scuti are dominated by a few low-degree ($l\lessapprox 3$) multiplets. \citet{aerts2010asteroseismology} offers a detailed explanation of spherical harmonics and multiplets.  \citet{burke2011effects} previously demonstrated the validity of perturbative methods to modelling acoustic mode oscillations in rotating $\delta$-Scuti stars, and \citet{ballot2010gravity} for gravity modes. 

For a non-rotating star, all $2l+1$ modes in a multiplet oscillate at the same frequency \citep{aerts2010asteroseismology}. For example, the $l=1$ multiplet contains a standing $m=0$ wave and prograde $m=1$ and retrograde $m=-1$ waves that run longitudinally across the stellar surface. However, stellar rotation is an additive effect with the $m=1$ and $m=-1$ running waves, making the $m=1$ wave appear faster, and the $m=-1$ wave appear slower. Therefore, the $m\neq 0$ modes in a multiplet change in frequency, with the negative-order modes decreasing and the positive-order modes increasing. This effect is known as rotational splitting, because in the frequency power spectrum of a photometric dataset, multiplet modes appear split apart due to stellar rotation. \citet{suarez2010use} showed that rotational splitting in rapidly rotating stars can be used to probe their internal rotation profile.

The amount a multiplet splits apart depends on how fast a star is rotating, with faster rotation resulting in larger $m\neq0$ frequency shifts. Therefore, multiplets inherently contain information about a star's rotation rate. For a slow rotator such as our Sun, $m=1,-1$ frequencies shift by tenths of a \mhz. However, the rapid rotation that commonly occurs in high-mass stars can result in frequency shifts of several tens of \mhz.   

Rapid stellar rotation fundamentally changes the physics of stellar oscillations. Multiplets in slow rotators are in equipartition, meaning all $2l+1$ modes oscillate at the same amplitude \citep[e.g.,][]{2019arXiv190202057K}. Therefore, the relative observed amplitude of the $m=0$ mode and the $m\neq0$ modes depends only on the star's obliquity angle. This property has been used to constrain the orbit geometries of dozens of transiting exoplanets \citep[e.g.,][]{huber2013stellar,chaplin2013asteroseismic,van2014asteroseismology,campante2016spin}. However, the assumption of equipartition is not valid in the regime of rapid rotation. Therefore, we unfortunately cannot reliably measure stellar obliquity angles of rapid rotators by comparing the relative amplitudes in a frequency power spectrum.

Fortunately, rotational splitting offers an independent approach for constraining stellar obliquities. Following \citet{goupil2000rotational} we apply second-order rotational splitting to measure KOI-976's rotation rate, 
\begin{equation}
\nu_m = \nu_0 + m(1-C_{\mathrm{nl}})\nu_\star + m^2D_1\frac{\nu_\star^2}{\nu_0}
\label{eq:rotsplit}
\end{equation}

where $\nu_i$ is the $m=i$ angular frequency in a multiplet, $\nu_\star$ is the stellar rotation frequency, and the Ledoux Constant $C_\mathrm{nl}$ and the second-order splitting constant and $D_1$ are values that depend on stellar properties \citep{goupil2000rotational}. $\delta$-Scuti stars often rotate near their breakup speed, so we include a second-order term that accounts for Coriolis and centrifugal force effects that can force rotational splitting to occur asymmetrically between prograde and retrograde modes.

To constrain KOI-976's rotation frequency, We start with Equation \ref{eq:rotsplit} and rearrange into two components by combining $m=(1,-1)$ solutions:
\begin{equation}
\frac{\nu_1-\nu_{-1}}{2}=(1-C_{\mathrm{nl}})\nu_\star
\label{eq:firstorder}
\end{equation}
and
\begin{equation}
\frac{\nu_1+\nu_{-1}-2\nu_0}{2} = \frac{D_1\nu_\star^2}{\nu_0}
\label{eq:secondorder}
\end{equation}

Equations \ref{eq:firstorder} and \ref{eq:secondorder} each affect a multiplet in different ways. The first-order term of Equation \ref{eq:firstorder} determines the total magnitude of frequency shifts. The second-order term of Equation \ref{eq:secondorder} affects the symmetry of the shift. For Sun-like stars, Equation \ref{eq:secondorder} is negligible due to the star's low rotation frequency. \citet{goupil2000rotational} notes rotational velocities $\gtrapprox100~\mathrm{km/s}$ require a third-order term in order to accurately model rotational splitting. However, we only apply second-order theory in this work, which increases the uncertainty in our determination of KOI-976's rotational frequency by $\sim0.14\mathrm{\mu Hz}$. Robust modelling of rapid stellar rotation is accessible through the 2D-code {\tt TOP} \citep[e.g.,][]{lignieres2006acoustic,reese2006acoustic}, the Adiabatic Code of Oscillation Including Rotation ({\tt ACOR}) \citep{ouazzani2012pulsations},  or the implicit two-dimensional hydrodynamic and hydrostatic code \citep{deupree1990stellar,deupree2010rotational}.

We use these equations to constrain the stellar rotation frequency, and then determine the stellar obliquity $\psi$ via, 

\begin{equation}
\psi=\cos^{-1}\left( \frac{v\sin(i)}{2\pi\nu_\star R_\star}\right)
\label{eq:obliquity}
\end{equation} 

We obtain the projected rotational velocity $v\sin(i)$ from \emph{Kepler's} Exoplanet Follow-Up Observing Program (ExoFOP) and list the value in Table \ref{table:parameters}. We detail the results of this analysis in \S\ref{sec:asteroresults}.

\subsubsection{Variability Subtraction}\label{sec:variability}
KOI-976's \emph{Kepler} photometry displays a significantly variable signal. The $\delta$-Scuti primary star oscillates with a signal that changes photometrically by $\sim$0.6\% with a fundamental oscillation period of $\sim$ 1.1 hours. These oscillations appear clearly in the short-cadence dataset; however, the long-cadence dataset's low sample rate does not properly resolve variability -- in fact, a cursory examination of the long-cadence photometry would lead one to believe that the dataset simply has a low signal-to-noise ratio. 

Rather than treating the primary star's variable signal as noise, we explicitly subtract KOI-976's modes of oscillation from short-cadence photometry in order to ``clean'' the transit light curve for detailed analysis. We consider this method a superior approach over phase-folding and binning the light curve for two reasons. First, phase-folding and binning can be risky when dealing with variability, particularly for oscillation frequencies near integer multiples of either the phase-folding frequency or the dataset's original sample frequency. And second, phase-folding and binning to remove stellar variability only works if a dataset is sufficiently large to properly average-out variability. In the specific case of KOI-976, the long-cadence dataset is $\sim$27 times as long as it's phase-folding period, which we find to be adequate-length baseline for averaging out stellar variability. However, many exoplanet photometric datasets (such as from NASA's Transiting Exoplanet Survey Satellite (\emph{TESS}) or ground-based observations) contain only one or a few transits, so stellar variability cannot be averaged out.

We subtract stellar variability from the short-cadence dataset using the variability-fitting program {\tt LASR} (Linear Algorithm for Significance Reduction) \citep{ahlers2018lasr}. This tool, which we developed specifically for subtracting oscillations from $\delta$-Scuti, removes oscillations from photometry one mode at a time by fitting the oscillation frequency, amplitude, and phase $(f,a,p)$ using the mode's statistical significance in frequency space as a goodness-of-fit criterion. By minimizing a given peak in the Lomb-Scargle normalized periodogram of the data, {\tt LASR} finds the sinusoidal solution that best removes a given mode of oscillation from photometry.

The {\tt LASR} fitting tool makes one assumption: that the stellar oscillations present in the photometry are well-modeled as a linear combination of sine waves. For $\delta$-Scuti, this assumption is generally considered acceptable \citep[e.g.,][]{breger2011regularities}. The Q9 short-cadence \emph{Kepler} photometric dataset spans about 6 weeks; modes of oscillation in $\delta$-Scuti typically do not change noticeably over those timescales. Additionally, $\delta$-Scuti variables typically do not exhibit non-sinusoidal variability such as starspots or flares due to their weak external magnetic fields. Therefore, {\tt LASR} is well-suited to model KOI-976's variable signal.

We apply {\tt LASR} to \emph{Kepler's} Q9 photometry of KOI-976 in order to clean the transit of stellar variability. We mask out the transit, working with only the out-of-transit baseline flux. We subtract oscillations modes one at a time from the dataset in order of descending amplitude until we reach our statistical significance cutoff of amplitude $\leq0.1\sigma$, where $\sigma$ is the average one standard deviation uncertainty of our time bins. At that limit, we cannot confidently distinguish between modes of oscillation and noise, so we end our cleaning process there. We then apply our sinusoidal solution to KOI-976's variability to the single transit in Q9 short-cadence photometry and subtract off stellar variability.

Worth noting is that our subtraction of variability from the transit is technically an oversubtraction. During the transit event, the binary companion blocks some of the primary star's flux -- and therefore some of the variable signal -- effectively decreasing the amplitude of the variable signal seen by \emph{Kepler}. We apply out-of-transit values for the $(f,a,p)$ of oscillation modes, which do not account for the transit event. However, we estimate that this oversubtraction can affect the transit light curve at a \emph{maximum} flux value of 0.18 mmag (versus the transit depth of 30 mmag). We show in our results that our {\tt LASR} subtraction provides a ``cleaned'' transit light curve with a roughly equivalent signal quality as phase-folding and binning the long-cadence dataset, and that this approach produces vastly superior results over phase-folding long-cadence datasets with only a handful of available transits. 

\begin{figure}[tbhp]
\epsscale{1}
\includegraphics[width = 0.45\textwidth]{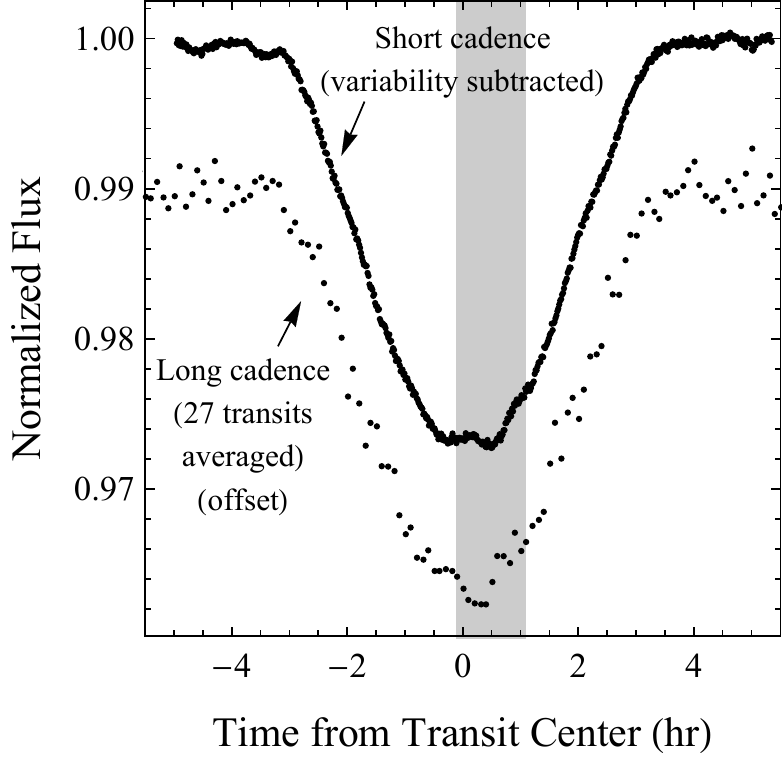}
\caption{\footnotesize KOI-976 transit light curves displaying the anomaly in their transits. We show the variability-subtracted short-cadence light curve on top and the long-cadence phase-folded  transit light curve binned at 480 seconds below. The gray area marks the unexplained sharp dip in brightness seen in all 27 long-cadence transit events and in the single short-cadence transit event. This repeating signal cannot be explained with our gravity-darkening model or with stellar variability, so we mask out the gray area during our transit analysis. We discuss possible causes of this anomaly in section \ref{sec:discanomaly}.}
\label{fig:anomaly}
\end{figure}

\subsection{Transit Light Curve Analysis} \label{sec:gravdark}
We analyze KOI-976's transit light curve with the gravity-darkening technique \citep{barnes2009transit}, which accounts for an oblate stellar shape and the pole-to-equator luminosity gradient induced by rapid stellar rotation. It takes advantage of the star's asymmetry in shape and luminosity to constrain the star's obliquity angle and the planet's projected alignment value directly from light curve fitting. \citet{ahlers2015spin} Figure 3 provides definitions of these angles.

We apply the gravity-darkening technique to both the long-cadence and short-cadence \emph{Kepler} time series to test the validity of variability subtraction as a method for preparing transit light curves. Following our previous gravity-darkening work \citep{2011ApJS..197...10B,ahlers2014,ahlers2015spin,barnes2015probable}, we use the light curve fitting package {\tt transitfitter}, which uses a Levenberg-Marqhardt $\chi^2$ minimization routine to model transit events across rapidly-rotating stars.

\renewcommand{\arraystretch}{1.2}
\begin{table*}
\centering
\begin{tabular}{l l l l l}
\hline \hline
{\bf Parameter} & {\bf Description} & {\bf Asteroseismology} & {\bf Short-Cadence Fit} & {\bf Long-Cadence Fit} \\ \hline
$\chi^2_\mathrm{red}$ & Goodness-of-fit & $3.81$ & $3.65$ & $1.94$ \\
$R_\star$ & Stellar radius $(R_\odot)$ & $--$ & $1.54\pm0.12$ & $1.59\pm0.15$ \\
$R_c$ & Companion radius $(R_\mathrm{Jup})$ & $--$ & $3.3\pm0.3$ & $3.9\pm0.4$ \\
$\Omega_\star$ & Stellar rotation rate $(\mathrm{hr})$ & $12.80\pm0.09$ & $14\pm3$ & $14\pm3$ \\
$\psi_\star$ & Stellar obliquity (deg) & $36^\circ\pm17^\circ$ & $46^\circ\pm16^\circ$ & $43^\circ\pm20^\circ$ \\
$\lambda$ & Projected alignment (deg) & $--$ & $7^\circ\pm13^\circ$ & $16^\circ\pm15^\circ$ \\
$i$ & Projected inclination (deg) & $--$ & $91.192^\circ\pm0.014^\circ$ & $91.28^\circ\pm0.03^\circ$ \\
$c_1$ & Limb-darkening constant & $--$ & $0.56(\pm0.2)$ & $0.56(\pm0.2)$ \\
$c_2$ & Limb-darkening constant & $--$ & $-0.16(\pm0.2)$ & $-0.16(\pm0.2)$\\
$\beta$ & Gravity-darkening exponent & $--$ & $0.17(\pm0.3)$ & $0.17(\pm0.3)$ \\
$\zeta$ & Stellar oblateness & $--$ & $0.045\pm0.007$ & $0.049\pm0.008$ \\
\end{tabular}
	\caption{\footnotesize Results from the combined asteroseismology and transit light curve analyses. Our rotational splitting measurements yield a high-precision measurement of the stellar rotation rate, but an imprecise constraint of the stellar obliquity because of our low-precision measurements of the star's radius. All of the transit fit results are lower-precision than expected because the companion star is in a grazing configuration, making the constraints of the star's radius and obliquity and the companion's transit geometry nearly degenerate. We apply assumed values for stellar limb-darkening, gravity-darkening, and the transiting object's eccentricity. We use assumed limb-darkening coefficients based on \citet{sing2010stellar}. We estimate KOI-976's gravity-darkening exponent based on previous gravity-darkening works \citep{2007Sci...317..342M,claret2011gravity}. Due to the grazing nature of the transiting companion, we assume a circular orbit and do not fit eccentricity. The $\chi^2_\mathrm{red}$ goodness-of-fit for our asteroseismology and short-cadence transit analyses are inflated due to the extremely small photometric uncertainty listed by \emph{Kepler} for the short-cadence dataset.}
\label{table:results}
\end{table*}

In our fit of the long-cadence dataset, we first adjust the entire time series according to the speed-of-light delay measured by \citet{balona2014binary} and described in \S \ref{sec:timedelay}. We then phase-fold the 27 transits in the time series around KOI-976's ExoFOP-reported orbit period (Table \ref{table:parameters}). We bin the resulting single transit light curve at 480 seconds to average out stellar variability and reduce computation time.

For the short-cadence dataset, we mask out the single transit in Q9 photometry and subtract off stellar variability based on our results from \S \ref{sec:variability}. We propagate our variability solution through the transit and subtract off all oscillations listed in Table 4.

We find a surprising but interesting anomaly in both the long-cadence and short-cadence transit light curves. Overall the light curves match the standard V-shape of a grazing binary. However, at about 60\% of the way through each transit event, a sharp drop in brightness occurs that lasts for roughly one hour and then goes away (Figure \ref{fig:anomaly}). This artifact appears in the same part of the transit in both variability-subtracted short-cadence data and phase-folded and binned long-cadence data, and so is not easily explained as a systematic in our datasets. The most straightforward explanation is that the dip in brightness is caused by a repeatable astrophysical phenomenon. Stellar gravity-darkening cannot produce this distinct drop in brightness, so we consider its analysis to be outside the scope of this paper. We mask out the unexplained signal when fitting the long-cadence and short-cadence transit light curves, and we discuss possible causes in \S\ref{sec:discanomaly}.

With the short-cadence and long-cadence time series fully prepared for fitting, we apply the gravity-darkening model to the transit light curves. We independently fit long-cadence and short-cadence data to contrast phase-folding and variability subtraction as approaches for handling transit light curves with stellar variability. KOI-976's companion transits in a grazing configuration; therefore, we cannot confidently extract information about its eccentricity \citep{barnes2007effects} and assume a circular orbit in our best-fit model. We show the results of our transit light curve analysis in \S\ref{sec:fitresults}.

\section{Results}

This work includes a four-stage analysis of KOI-976, a rapidly-rotating $\delta$-Scuti with a transiting binary companion observed by \emph{Kepler}. We first measure the primary star's obliquity angle and rotation rate by applying asteroseismic theory to KOI-976's seismically-active out-of-transit photometry. We then subtract out stellar variability from KOI-976's short-cadence transit event to prepare the dataset for further analysis. Finally, we perform a joint analysis of both the long-cadence and short-cadence \emph{Kepler} transit light curves using a gravity-darkened model to measure bulk parameters of the system and to test the validity of variability subtraction as a path for cleaning transit light curves. We describe our results in the following subsections.

\subsection{Asteroseismology}\label{sec:splitresults}
\begin{figure*}[tbhp]
\epsscale{1}
\includegraphics[trim=0 0 0 0, clip, width=\textwidth]{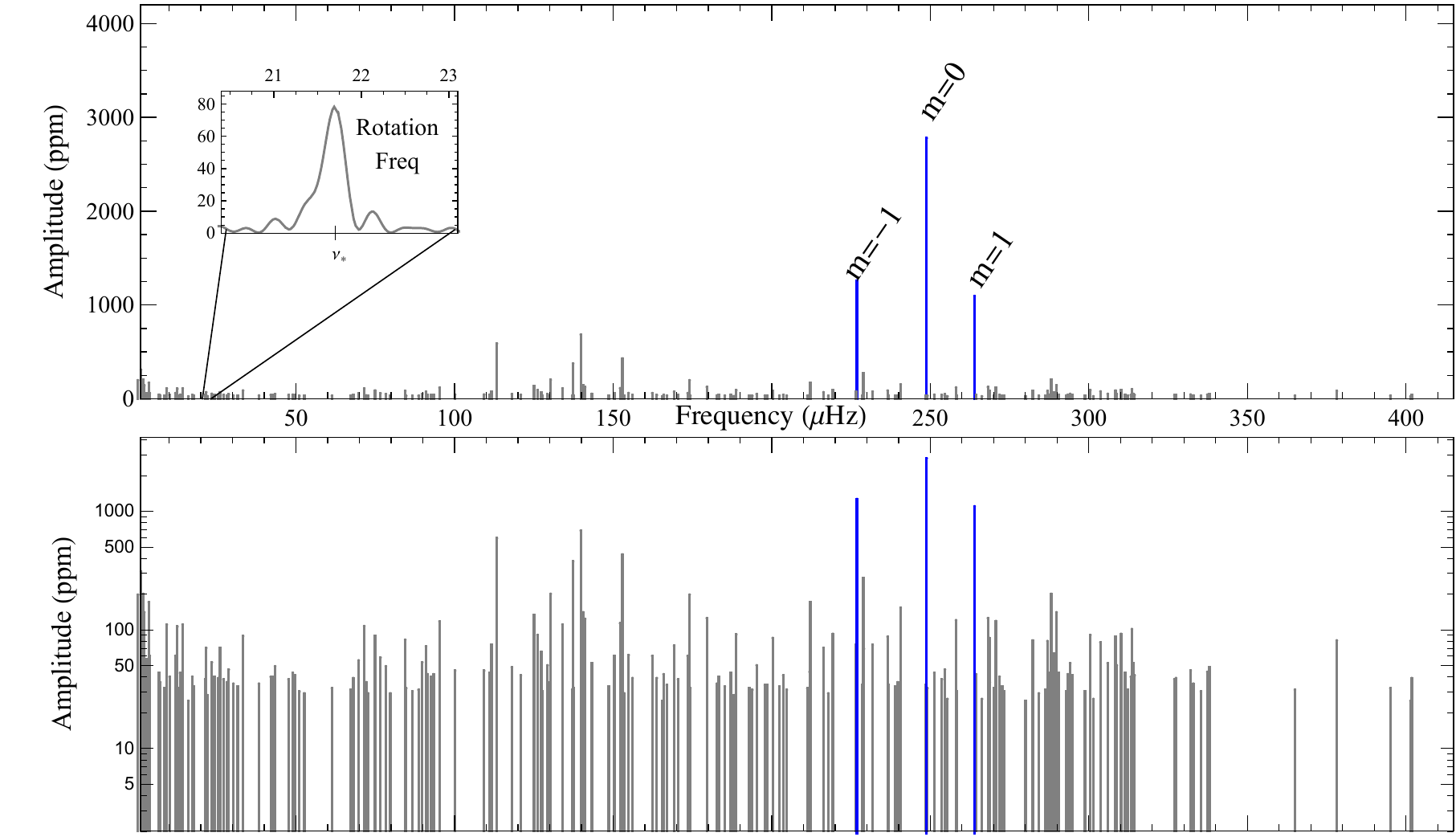}
\caption{\footnotesize (Top) KOI-976's frequency power spectrum. We identify a frequency triplet (shown in blue), with azimuthal modes $(m)$ labelled. The triplet displays large-scale and asymmetric rotational splitting, consistent with rapid stellar rotation. We color the triplet's corresponding measured frequencies blue in Table \ref{table:oscillations}. The lower-frequency dominant modes in KOI-976's power spectrum do not match rotational splitting theory \citep{dziembowski1992effects,soufi1998effects}. (Bottom) Logarithmic plot showing KOI-976's 204 identified modes of oscillation.}
\label{fig:periodogram}
\end{figure*}

We measure KOI-976's rotation rate using second-order rotational splitting. We find the $(\nu_{-1},\nu_0,\nu_1)$ frequencies of the dominant triplet through inspection of KOI-976's power spectrum, shown in Figure \ref{fig:periodogram} and listed in Table 4. We set an initial range of possible $\nu_\star$ values by applying the range $0.8\leq (1-C_{nl})\leq1.0$ for the Ledoux constant, which encompasses both the theoretical and emperical ranges of values expected for $\delta$-Sct stars. This initial range constrains the star's rotation frequency to $18.5\mathrm{\mu Hz}\leq\nu_\star\leq23.1\mathrm{\mu Hz}$ through Equation \ref{eq:firstorder}. We list our rotational splitting results in Table \ref{table:splitresults}.

\renewcommand{\arraystretch}{1.2}
\begin{table}[h] 
\centering
\begin{tabular}{lr}
\hline
\hline
{\bf Parameter} & {\bf Value} \\ \hline
$C_{nl}$ & $0.08776\pm0.00003$ \\
$D_1$ & $1.97597\pm0.00013$ \\
$\nu_\star(\mathrm{\mu Hz})$ & $21.71\pm0.16$ \\
$v\sin(i)(\mathrm{km/s})$ & $120\pm5$ \\
\end{tabular}
\caption{Measured parameters from rotational splitting. We color-code the rotational splitting constants according to their multiplet, matching Figure \ref{fig:periodogram}. The Ledoux constant falls within the expected range for typical $\delta$-Scuti stars. We combine these results and our constraints obtained from transit light curve fitting to measure the star's obliquity angle, which we list in Table \ref{table:results}.}
\label{table:splitresults}
\end{table}

Only three frequency peaks exist in the possible range of rotation frequencies. Following \citet{breger2011regularities}, we identify the rotation frequency peak to be 21.705~\mhz~based on the peak's slightly asymmetric shape and significance (Figure \ref{fig:periodogram} inset). Our identified $\nu_\star$ acts as the combination frequency for most of the modes listed in Table \ref{table:oscillations}, further suggesting that the identified peak is the star's rotation frequency. We identify the combination frequencies as a check for identifying KOI-976's rotation frequency and do not consider the combinations listed in Table \ref{table:oscillations} to be an exhaustive search.

We constrain the primary star's obliquity angle using Equation \ref{eq:obliquity}. We use the emperical mass-radius relation \citep{demircan1991stellar} to set an approximate range for the stellar radius: $1.48R_\odot\leq R_\star\leq1.85R_\odot$. This range of stellar radii agrees with the short-cadence and long-cadence transit light curve fit results (see Table \ref{table:results}).

\subsection{Variability Subtraction} \label{sec:asteroresults}
\begin{figure*}[tbhp]
\centering
\begin{tabular}{r l}
\vspace*{-1cm} \\
\includegraphics[width=0.48\textwidth]{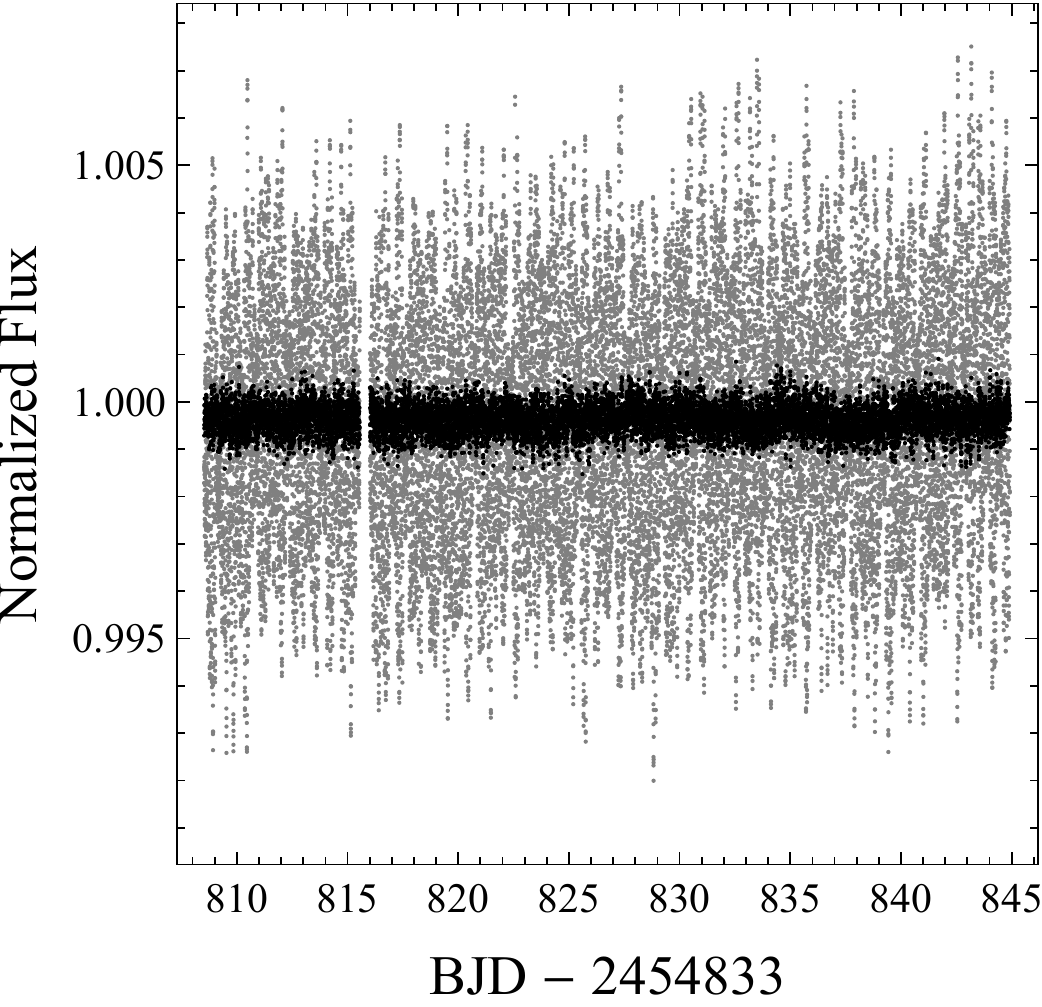} & \hspace*{-0.26cm}\includegraphics[width=0.48\textwidth]{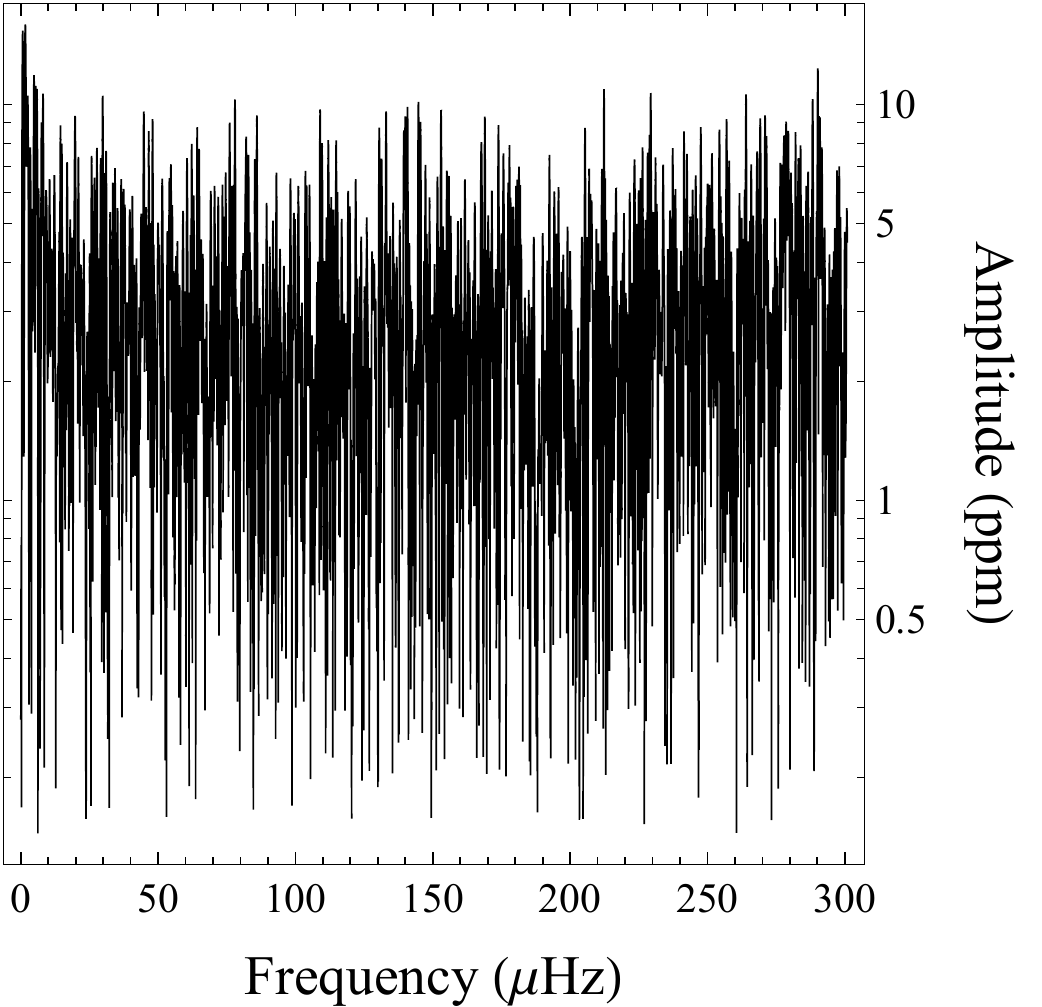}\\
\end{tabular}
\caption{\footnotesize (Left) Original (gray) and subtracted (black) Q9 short-cadence photometry. We subtract off the 204 frequencies listed in Table \ref{table:oscillations} to clean the time series and prepare it for transit light curve analysis. The data gap near 815 days is the transit event, which we mask out of the dataset during the variability subtraction process. We extend our subtraction results through the transit event to remove variability (see Figure \ref{fig:anomaly}.) (Right) KOI-976's frequency power spectrum after subtracting all identified modes of oscillation.}
\label{fig:scbaseline}
\end{figure*}

We use the programming tool {\tt LASR} \citep{ahlers2018lasr} to subtract off 204 oscillation modes from KOI-976's Q9 short-cadence \emph{Kepler} photometry. We find the $\delta$-Scuti's variability to be well-modeled as a linear combination of sinusoids over the $\sim$ six week dataset. We list the best-fit frequencies, amplitudes, and phases of all oscillation modes in Table 4 and show the subtraction results in Figure \ref{fig:scbaseline}.

\subsection{Light Curve Fits}\label{sec:fitresults}

\begin{figure*}[tbhp]
\begin{tabular}{r l}
\vspace*{-1cm} \\
\includegraphics[width=0.49\textwidth]{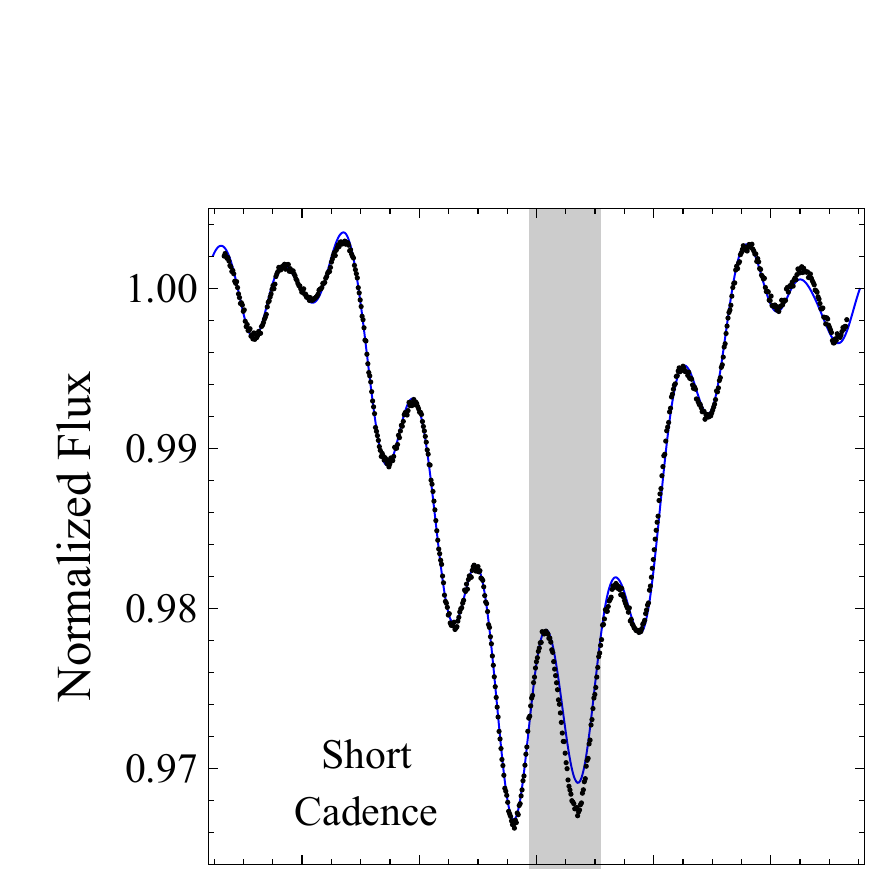} & \hspace*{-0.37cm}\includegraphics[width=0.49\textwidth]{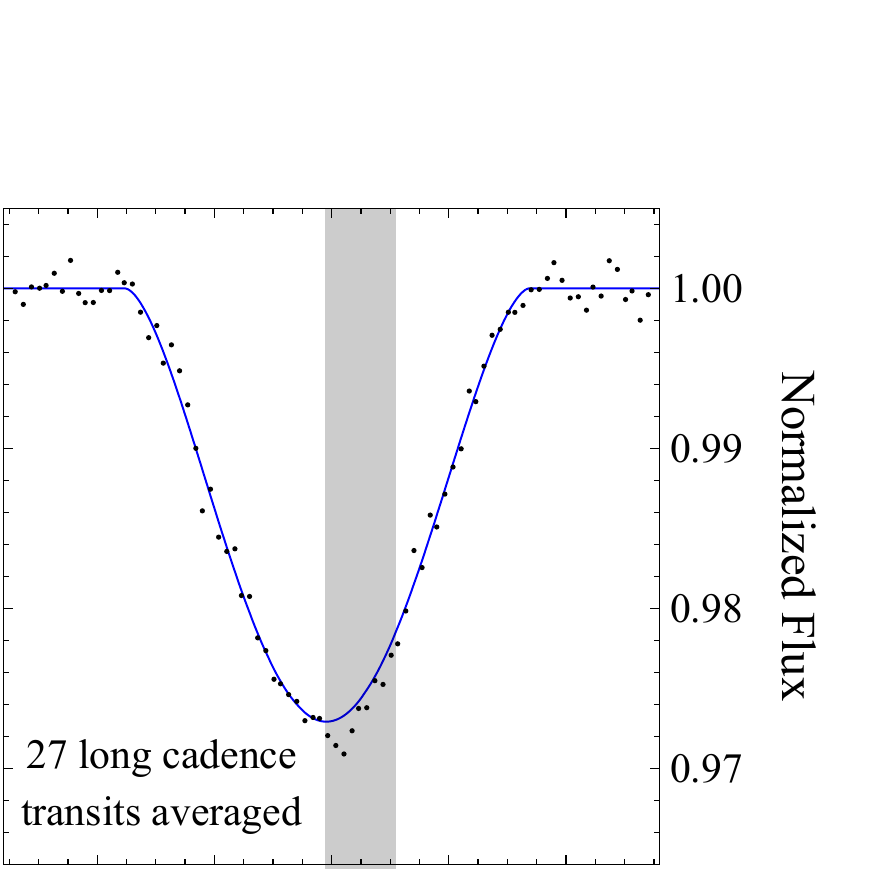}\\
\vspace*{-0.58cm}\\
\includegraphics[width=0.49\textwidth]{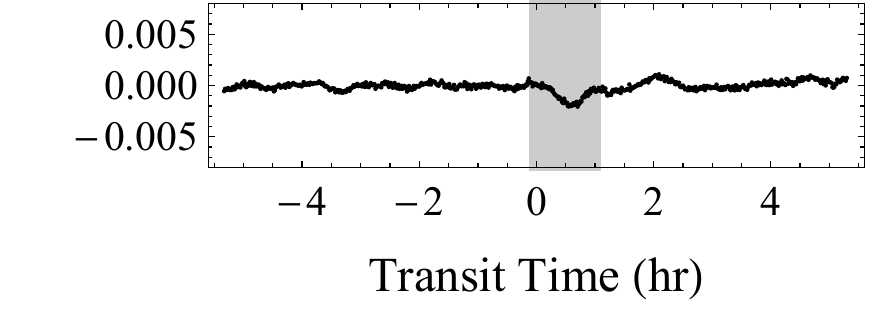} & \hspace*{-0.37cm}\includegraphics[width=0.49\textwidth]{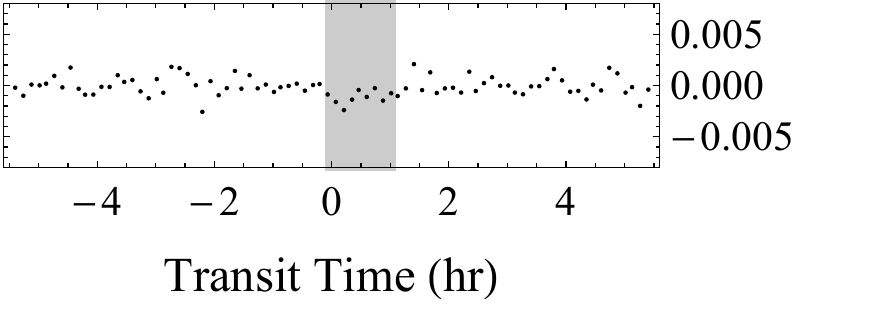} 
\end{tabular}
\caption{\footnotesize KOI-976 short-cadence (left) and long-cadence (right) \emph{Kepler} transit light curves and best-fit models. The short-cadence best-fit model includes both the gravity-darkened fit and the 204 oscillation modes resolved in the dataset. The long-cadence light curve includes 27 phase-folded transits and is binned to average out stellar variability. The gaps in the datasets and their residuals correspond to masking out the transit analomaly (see Figure \ref{fig:anomaly}). The short-cadence approach of variability subtraction yields a slightly better overall fit than averaging out variability through phase-folding with residual standard deviations of $\sigma_\mathrm{SC}=3\times10^{-4}$ and $\sigma_\mathrm{LC}=9\times10^{-4}$.}
\label{fig:bestfit}
\end{figure*}

After subtracting stellar variability from KOI-976's short-cadence light curve and masking out the anomalous bump seen in Figure \ref{fig:anomaly}, we fit the resulting transit using the gravity-darkening technique. We show the best-fit and residual in Figure \ref{fig:bestfit} and list the results in Table \ref{table:results}.

We also phase-fold KOI-976's long-cadence dataset around the ExoFOP-reported orbital period of $52.56902\pm5\times10^{-5}$ days and bin the light curve at 480 seconds to average out stellar variability. We fit both datasets with the gravity-darkening model to obtain contrains on stellar radius, companion radius, stellar obliquity, and the companion's projected alignment and inclination angles. For both fits, we use quadratic limb-darkening coefficient values from \citet{sing2010stellar} and ascribe uncertainties to them based on the star's temperature uncertainty.

\begin{figure}[tbhp]
\centering
\includegraphics[width=0.37\textwidth]{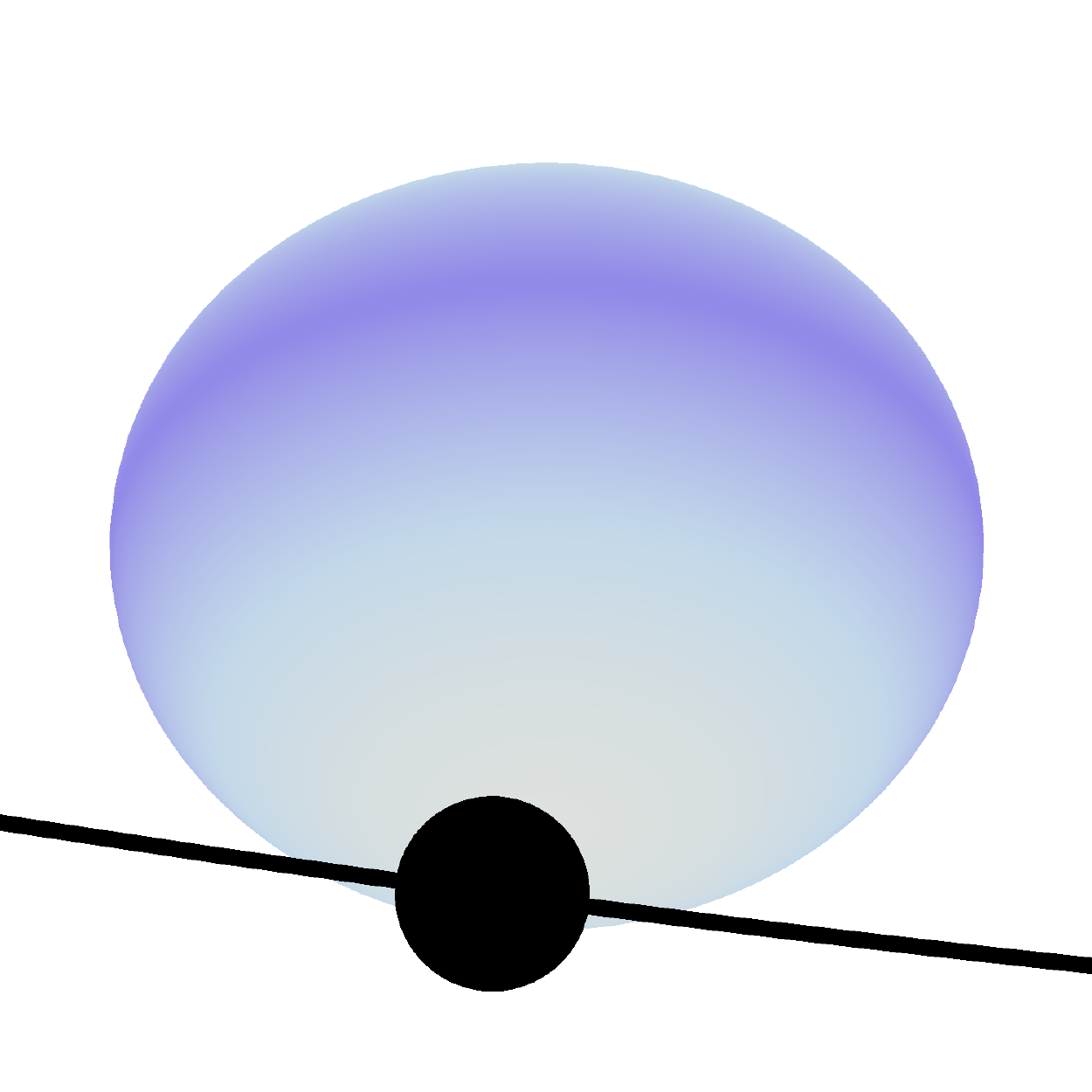}
\caption{\footnotesize Our best-estimate of KOI-976's transit geometry. We find that the companion star transits near one of the primary star's brighter poles in a grazing configuration. Our transit  our results contain a high uncertainty for the transit geometry due to its grazing nature, but asteroseismic analysis provides a reliable obliquity value for the primary star.}
\label{fig:transitgeometry}
\end{figure}

We find that KOI-976 is a binary system in a grazing configuration with no secondary eclipse, consistent with previous observations. The \emph{Kepler} telescope provided very high-quality photometry of this system, normally allowing for detailed analysis of the transit light curve; however, the grazing nature of the transit introduces strong interdependencies in our fitting model between the star's radius and projected orbital inclination. The projected inclination (closely related to the transit impact parameter) describes how much of the companion actually passes in front of the the star, directly affecting both the transit duration and the transit depth. This effect brings about the large uncertainties seen in Table \ref{table:results}. We display our best estimate of the transit geometry in Figure \ref{fig:transitgeometry}.

Even with all 27 long-cadence transits folded together, KOI-976's single variability-subtracted short-cadence transit provides three times finer precision than its long-cadence counterpart. The standard deviations of the residuals shown in Figure \ref{fig:bestfit} are $\sigma_\mathrm{SC}=3\times10^{-4}$ and $\sigma_\mathrm{LC}=9\times10^{-4}$ for the short-cadence and long-cadence best-fit models, respectively, implying that variability subtraction provides a robust path for cleaning transit light curves of stellar seismic activity.

\section{Discussion} \label{sec:discussion}
This work tests the feasibility of a combined asteroseismic and transit light curve analysis of planets transiting high-mass variable stars. NASA's \emph{TESS} mission will likely discover over a thousand exoplanets orbiting rapidly-rotating stars, and as many as a few hundred of them will orbit $\delta$-Scuti or related variable stars \citep{barclay2018revised}. We demonstrate that the obstacles of rapid rotation and variability can be overcome and even used advantageously during transit light curve analysis. 

Asteroseismology can provide a wealth of information about host stars. We only skim the surface of asteroseismology's capabilities in this work by applying rotational splitting theory to measure the host star's external rotation rate and obliquity angle. $\delta$-Scuti possess the fortuitous characteristic of typically being dominated by a few low-order modes of oscillation, making rotational splitting an accessible form of analysis. We identify a frequency triplet in Figure \ref{fig:periodogram} that dominates the variability seen in KOI-976's \emph{Kepler} photometry. We find that the triplet displays asymmetric rotational splitting consistent with rapid stellar rotation and identify the star's rotation in its frequency power spectrum following previous works on $\delta$-Scuti \citep[e.g.,][]{breger2000delta,breger2011regularities}.

Our results demonstrate that variability subtraction for $\delta$-Scuti and related variable stars provides a reliable process for making transit light curve analysis possible. Using the variability-removal program {\tt LASR} \citep{ahlers2018lasr}, we show that the single short-cadence transit event can provide superior photometric precision to removing variability by phase-folding. 

We find that this system resides in a spin-orbit misaligned configuration. Our asteroseismic and light curve fitting procedures measure the host star's obliquity rate to be $42^\circ\pm10^\circ$ when treating the three approaches as three independent measurements. Exoplanets (and binary companions) orbiting high-mass stars ($M_\star\geq1.3M_\odot$) are expected to reside in misaligned orbits far more commonly than for those orbiting low-mass stars, so our results are consistent with other works on spin-orbit misalignment \citep{winn2010hot,albrecht2012obliquities,dawson2014tidal}. 

\subsection{$\delta$-Scuti Stars and \emph{TESS}}
We choose KOI-976 as our example system for this work because this star is a prototypical rapidly-rotating $\delta$-Scuti and because of its exceptional signal-to-noise. However, \emph{TESS} photometry will not achieve the same precision as \emph{Kepler}; with an effective aperture size of 10 cm \citep{ricker2014transiting}, \emph{TESS} will only obtain approximately one hundredth the precision of its predecessor. Therefore future transit light curve analyses of rapidly-rotating $\delta$-Scuti will likely not involve such detailed resolution of stellar variability or of transit events. 

The gravity-darkening technique will still apply for low-signal photometry. \citet{barnes2015probable} demonstrated that gravity-darkening can still make constraints on a planet's orbit geometry even when a transit event is barely visible. Similarly, gravity-darkening can provide upper limits on the projected alignment of \emph{TESS} objects orbiting rapid rotators based on the gravity-darkening-induced asymmetry (or lack thereof) in their transit light curves.

Detection of rotational splitting is also possible from low-precision photometry. $\delta$-Scuti stars are typically dominated by a few low-order modes, making multiplets often easy to identify in frequency space. For rapid rotators, multiplets split apart by several tens of \mhz, making them easy to resolve in frequency space even with low-precision data and with only 27 days of baseline photometry. \citet{ahlers2018lasr} shows that oscillations can be reliably measured when their amplitudes are as little as one-tenth of the data's noise level. 

Variability subtraction will be particularly useful for \emph{TESS} systems. \S\ref{sec:fitresults} shows that variability subtraction yields a better result than phase-folding transits even with 27 available transit events to average out the seismic signal. Most \emph{TESS} systems will only have one or a few transits available, so phase-folding to average out variability such as KOI-976's will not be viable. Variability subtraction will be the most viable method of overcoming stellar variability during the \emph{TESS} era.

\subsection{Transit Anomaly}\label{sec:discanomaly}
The transit anomaly shown in Figure \ref{fig:anomaly} is a sharp, $\sim1$ hour drop in brightness approximately 60\% of the way through the transit. This signal appears in the short-cadence transit light curve after stellar variability has been subtracted and appears in each of KOI-976's 27 individual long-cadence transits. The dimming event also appears in the phase-folded version of the long-cadence transit. We therefore suspect that the anomaly is astrophysical and not a systematic in our dataset because the event occurs periodically in the same part of KOI-976's transit events with the same basic shape, duration, and depth, and appears in both short-cadence and long-cadence photometry. 

Detailed analysis of the unusual signal is outside the scope of this work; however, we rule out many possibilities through simple thought exercises. Many events can cause a decrease in photometric brightness, but no straightforward phenomena match the observed signal.

The existence of an additional transiting body likely cannot explain the short, sudden decrease in brightness shown in Figure \ref{fig:anomaly}. The anomaly occurs every transit, so an object with one-half the binary companion's orbit period could cause a signal in sync with every primary transit event. The anomaly's signal in variability-subtracted short-cadence photometry is $2\sigma$ larger than the out-of-transit baseline noise, but we find no evidence of the anomaly occurring outside of the transit event.

Rings or an accretion disk around the transiting body also cannot produce the anomaly. \citet{barnes2004transit} and others \citep{kenworthy2015modeling,des2017search,hatchett2018pilot,aizawa2018systematic} model the possible phtometric signals a ring system or accretion disk can produce during a transit/eclipse. Most prominent in transits with rings is an increased effective radius of the transiting body and disctint jumps upward in the light curve due to light passing through gaps in the rings. Rings/accretion disks do not match the anomalous signal in Figure \ref{fig:anomaly} and can be ruled out.

The anomaly displays the exact opposite characteristics of a transit across a starspot. Rather than a short-lived increase in photometric signal as seen with stroboscopic starspots \citep[e.g.,][]{desert2011hot,sanchis2011starspots}, we see a distinct drop in brightness as if the transiting companion were passing in front of a local hot spot on the primary star. Stellar hot spots have been observed in the past that were likely brought about by instabilities in the star's magnetic field, but only for T Tauri stars \citep{kenyon1994hot}. No prior observations indicate that a main-sequence star variable can possess local hot spots on their surface. The anomaly appears at the same time in every transit event, so if it is caused by a hot spot or some other form of local instability and must not be affected by the star's rotation rate. The hot spot would therefore need to be located at one of the host star's poles, which is consistent with our transit geometry. However, validation of the hot spot hypothesis for this transit signal is outside the scope of this work.

\section{Conclusion}
The gravity-darkening, rotational splitting, and variability subtraction techniques described in this manuscript provide new windows for studying planets and binary companions orbiting high-mass stars. A/F-type stars commonly exhibit rapid rotation and stellar variability that can obfuscate detailed analysis of transit photometry. We demonstrate how to overcome these challenges and use them to constrain the transiting body's orbit geometry.

Exoplanets orbiting high-mass stars such as KOI-976 commonly reside in spin-orbit misaligned positions \citep{winn2010hot}. While the underlying mechanisms for causing misalignment are still under investigation, recent observations have found many high-mass stars to host severely inclined or even retrograde planets \citep[e.g.,][]{2011ApJS..197...10B,2011AJ....141...63W,2012ApJ...757...18A,ahlers2015spin,gaudi2017giant}. 

With better constraints on the distribution of alignment angles around high-mass stars, the dominant mechanisms for causing exoplanets to misalign will become clearer. Measuring bulk parameters and orbit geometries of systems such as KOI-976 helps constrain the formation and evolution pathways of such systems and helps explain the apparent dichotomy between high-mass and low-mass system geometries.

This work demonstrates one approach to measuring system parameters around active, rapidly-rotating high-mass stars that applies well to NASA's \emph{TESS} mission. With a spectroscopically-determined $v\sin(i)$ value for the host star, one can potentially obtain two independent measurements of a transiting body's orbit geometry directly from \emph{TESS} photometry using asteroseismology and gravity-darkening. The techniques detailed in this manuscript will likely apply to as many as several hundred \emph{TESS} systems in the near future.

\bibliographystyle{apj}
\bibliography{citations}

\clearpage

\renewcommand{\arraystretch}{1.15}
\begin{table}
\label{table:oscillations}
\begin{tabular}{lrrrl}
\hline 
\hline
{\bf \#} & {\bf Freq $(\mathrm{\mu Hz})$} & {\bf Amp$(10^{-3})$} & {\bf Phase} \\ \hline
{\color{blue}$1$}	&	{\color{blue}$248.8552\pm0.0002$}	&	{\color{blue}$2.778\pm0.005$}	&	{\color{blue}$1.614\pm0.003$}	& {\color{blue}$\nu_1$} \\
{\color{blue}$2$}	&	{\color{blue}$226.8755\pm0.0004$}	&	{\color{blue}$1.261\pm0.010$}	&	{\color{blue}$6.049\pm0.005$}	&	{\color{blue}$\nu_2$}	\\
{\color{blue}$3$}	&	{\color{blue}$263.9633\pm0.0018$}	&	{\color{blue}$1.094\pm0.004$}	&	{\color{blue}$4.86\pm0.02$}	&	{\color{blue}$\nu_3$}	\\
$4$	&	$139.8156\pm0.0013$	&	$0.684\pm0.006$	&	$3.72\pm0.03$	&	$\nu_4$	\\
$5$	&	$113.3531\pm0.0013$	&	$0.589\pm0.006$	&	$5.046\pm0.019$	&	$\nu_5$	\\
$6$	&	$152.961\pm0.005$	&	$0.444\pm0.010$	&	$2.45\pm0.06$	&	$\nu_6$	\\
$7$	&	$137.3857\pm0.0013$	&	$0.375\pm0.008$	&	$6.116\pm0.016$	\\
$8$	&	$288.132\pm0.008$	&	$0.2\pm0.02$	&	$3.58\pm0.08$	&	$2\nu_3-\nu_6-4\nu_\star$	\\
$9$	&	$228.914\pm0.006$	&	$0.27\pm0.018$	&	$4.71\pm0.08$	\\
$10$	&	$130.326\pm0.005$	&	$0.209\pm0.017$	&	$6.00\pm0.04$	&	$-\nu_4+2\nu_5+2\nu_\star$	\\
$11$	&	$174.129\pm0.005$	&	$0.197\pm0.005$	&	$1.19\pm0.04$	&	$2\nu_2-2\nu_4$	\\
$12$	&	$212.200\pm0.009$	&	$0.17\pm0.02$	&	$1.97\pm0.05$	\\
$13$	&	$240.661\pm0.005$	&	$0.15\pm0.01$	&	$2.76\pm0.04$	\\
$14$	&	$140.604\pm0.016$	&	$0.148\pm0.012$	&	$5.29\pm0.10$	&	$-\nu_2+2\nu_4+\nu_6-3\nu_\star$	\\
$15$	&	$289.802\pm0.004$	&	$0.142\pm0.017$	&	$3.14\pm0.04$	\\
$16$	&	$125.073\pm0.006$	&	$0.133\pm0.005$	&	$3.87\pm0.08$	&	$\nu_3-2\nu_5+\nu_6-3\nu_\star$	\\
$17$	&	$179.569\pm0.009$	&	$0.125\pm0.008$	&	$0.34\pm0.07$	&	$\nu_5+\nu_6-4\nu_\star$	\\
$18$	&	$268.192\pm0.003$	&	$0.125\pm0.004$	&	$4.02\pm0.03$	&	$\nu_1+\nu_3-\nu_5-\nu_6+\nu_\star$	\\
$19$	&	$258.122\pm0.004$	&	$0.120\pm0.007$	&	$2.1\pm0.04$	\\
$20$	&	$134.051\pm0.005$	&	$0.121\pm0.01$	&	$3.81\pm0.05$	\\
$21$	&	$270.701\pm0.006$	&	$0.118\pm0.009$	&	$0.17\pm0.07$	\\
$22$	&	$152.206\pm0.010$	&	$0.113\pm0.006$	&	$4.9\pm0.1$	&	$\nu_2-\nu_4+3\nu_\star$	\\
$23$	&	$95.283\pm0.006$	&	$0.117\pm0.009$	&	$5.57\pm0.08$	&	$-2\nu_1+2\nu_3+3\nu_\star$	\\
$24$	&	$313.581\pm0.006$	&	$0.101\pm0.006$	&	$5.77\pm0.07$	&	$2\nu_5+4\nu_\star$	\\
$25$	&	$71.53\pm0.007$	&	$0.107\pm0.007$	&	$3.47\pm0.07$	\\
$26$	&	$268.626\pm0.01$	&	$0.085\pm0.009$	&	$0.62\pm0.10$	\\
$27$	&	$300.480\pm0.007$	&	$0.093\pm0.010$	&	$3.54\pm0.07$	&	$\nu_2+\nu_4-\nu_6+4\nu_\star$	\\
$28$	&	$188.831\pm0.01$	&	$0.091\pm0.007$	&	$3\pm0.11$	\\
$29$	&	$236.550\pm0.005$	&	$0.087\pm0.005$	&	$1.96\pm0.06$	&	$\nu_1-\nu_2+2\nu_4-3\nu_\star$	\\
$30$	&	$219.271\pm0.010$	&	$0.092\pm0.006$	&	$1.16\pm0.06$	\\
$31$	&	$310.155\pm0.009$	&	$0.092\pm0.008$	&	$5.96\pm0.08$	&	$\nu_1-\nu_5+\nu_6+\nu_\star$	\\
$32$	&	$126.191\pm0.007$	&	$0.088\pm0.010$	&	$4.74\pm0.06$	&	$\nu_1+\nu_4-2\nu_6+2\nu_\star$	\\
$33$	&	$308.423\pm0.006$	&	$0.08\pm0.01$	&	$1.78\pm0.06$	&	$2\nu_1-\nu_3+\nu_4-3\nu_\star$	\\
$34$	&	$141.167\pm0.007$	&	$0.123\pm0.008$	&	$3.64\pm0.07$	&	$-\nu_3+3\nu_5+3\nu_\star$	\\
$35$	&	$74.95\pm0.01$	&	$0.08\pm0.01$	&	$2.95\pm0.06$	&	$-\nu_1-\nu_5+3\nu_6-\nu_\star$	\\
$36$	&	$200.328\pm0.009$	&	$0.08\pm0.01$	&	$0.36\pm0.10$	&	$2\nu_3-2\nu_6-\nu_\star$	\\
$37$	&	$282.288\pm0.016$	&	$0.08\pm0.01$	&	$5.52\pm0.10$	\\
$38$	&	$378.185\pm0.01$	&	$0.081\pm0.008$	&	$3.64\pm0.1$	\\
$39$	&	$84.462\pm0.006$	&	$0.082\pm0.008$	&	$0.82\pm0.09$	&	$\nu_3-\nu_5-\nu_6+4\nu_\star$	\\
$40$	&	$303.656\pm0.009$	&	$0.078\pm0.007$	&	$5.81\pm0.09$	&	$2\nu_1-\nu_3+\nu_5-2\nu_\star$	\\
$41$	&	$497.713\pm0.008$	&	$0.077\pm0.008$	&	$4.6\pm0.1$	&	$2\nu_1$	\\
$42$	&	$286.949\pm0.008$	&	$0.08\pm0.006$	&	$5.11\pm0.09$	\\
$43$	&	$231.807\pm0.013$	&	$0.074\pm0.009$	&	$2.31\pm0.08$	&	$2\nu_2+\nu_4-3\nu_5-\nu_\star$	\\
$44$	&	$226.628\pm0.006$	&	$0.074\pm0.009$	&	$3.52\pm0.1$	&	$\nu_4+4\nu_\star$	\\
$45$	&	$111.685\pm0.007$	&	$0.074\pm0.008$	&	$1.3\pm0.1$	\\
\hline
\end{tabular}
\end{table}

\begin{table}
\begin{tabular}{lrrrl}
\hline
\hline
{\bf \#} & {\bf Freq $(\mathrm{\mu Hz})$} & {\bf Amp$(10^{-3})$} & {\bf Phase} \\ \hline
$46$	&	$91.013\pm0.013$	&	$0.072\pm0.007$	&	$0.1\pm0.08$	&	$2\nu_2-\nu_4-\nu_5-\nu_6+2\nu_\star$	\\
$47$	&	$169.236\pm0.017$	&	$0.073\pm0.009$	&	$4.52\pm0.11$	&	$\nu_2+2\nu_3-2\nu_4-2\nu_6$	\\
$48$	&	$33.32\pm0.01$	&	$0.088\pm0.008$	&	$5.77\pm0.08$	\\
$49$	&	$216.347\pm0.010$	&	$0.07\pm0.009$	&	$3.68\pm0.09$	&	$\nu_1-\nu_3+\nu_4+\nu_5-\nu_\star$	\\
$50$	&	$500.264\pm0.012$	&	$0.064\pm0.008$	&	$0\pm0.10$	\\
$51$	&	$12.608\pm0.005$	&	$0.106\pm0.009$	&	$5.48\pm0.06$	&	$2\nu_2-3\nu_4-\nu_\star$	\\
$52$	&	$14.284\pm0.010$	&	$0.117\pm0.013$	&	$3.93\pm0.07$	\\
$53$	&	$21.705\pm0.009$	&	$0.069\pm0.013$	&	$3.83\pm0.1$	&	$\nu_\star$	\\
$54$	&	$26.074\pm0.008$	&	$0.072\pm0.010$	&	$4.6\pm0.09$	&	$-\nu_1+3\nu_5-3\nu_\star$	\\
$55$	&	$76.653\pm0.009$	&	$0.058\pm0.009$	&	$2.62\pm0.1$	\\
$56$	&	$173.513\pm0.012$	&	$0.06\pm0.009$	&	$4.16\pm0.14$	\\
$57$	&	$150.403\pm0.013$	&	$0.06\pm0.006$	&	$3.91\pm0.11$	&	$\nu_3-3\nu_4+2\nu_6$	\\
$58$	&	$154.897\pm0.010$	&	$0.061\pm0.009$	&	$5.70\pm0.1$	&	$\nu_1+\nu_3-\nu_4-\nu_6-3\nu_\star$	\\
$59$	&	$127.347\pm0.010$	&	$0.065\pm0.008$	&	$1.96\pm0.10$	&	$\nu_4-3\nu_5+2\nu_6+\nu_\star$	\\
$60$	&	$127.62\pm0.02$	&	$0.03\pm0.007$	&	$1.4\pm0.3$	\\
$61$	&	$162.348\pm0.013$	&	$0.06\pm0.010$	&	$2.93\pm0.13$	&	$-\nu_2+2\nu_4+\nu_6-2\nu_\star$	\\
$62$	&	$69.754\pm0.010$	&	$0.055\pm0.006$	&	$4.83\pm0.14$	&	$2\nu_3-\nu_4-3\nu_5+\nu_\star$	\\
$63$	&	$43.429\pm0.016$	&	$0.049\pm0.007$	&	$5.53\pm0.17$	&	$2\nu_\star$	\\
$64$	&	$143.277\pm0.010$	&	$0.052\pm0.008$	&	$1.5\pm0.16$	\\
$65$	&	$89.819\pm0.010$	&	$0.053\pm0.008$	&	$3.04\pm0.17$	&	$-\nu_1+\nu_3+\nu_4-3\nu_\star$	\\
$66$	&	$23.487\pm0.014$	&	$0.053\pm0.009$	&	$1.09\pm0.13$	&	$2\nu_2-\nu_3-2\nu_4+\nu_5$	\\
$67$	&	$229.079\pm0.008$	&	$0.069\pm0.009$	&	$5.61\pm0.09$	&	$-\nu_1-2\nu_2+3\nu_3+\nu_4$	\\
$68$	&	$305.959\pm0.013$	&	$0.05\pm0.01$	&	$2.56\pm0.17$	&	$2\nu_6$	\\
$69$	&	$338.033\pm0.014$	&	$0.048\pm0.009$	&	$0.8\pm0.2$	&	$-\nu_1+3\nu_3-2\nu_5+\nu_\star$	\\
$70$	&	$118.085\pm0.016$	&	$0.048\pm0.009$	&	$5.87\pm0.17$	&	$\nu_4-\nu_\star$	\\
$71$	&	$313.983\pm0.017$	&	$0.052\pm0.008$	&	$1.93\pm0.11$	&	$\nu_1+3\nu_\star$	\\
$72$	&	$294.120\pm0.013$	&	$0.052\pm0.007$	&	$4.95\pm0.19$	\\
$73$	&	$78.362\pm0.017$	&	$0.049\pm0.008$	&	$1.70\pm0.14$	&	$-2\nu_4+2\nu_5+\nu_6-\nu_\star$	\\
$74$	&	$289.055\pm0.016$	&	$0.063\pm0.007$	&	$3.53\pm0.08$	\\
$75$	&	$308.809\pm0.010$	&	$0.05\pm0.007$	&	$5.65\pm0.17$	\\
$76$	&	$187.117\pm0.014$	&	$0.04\pm0.01$	&	$6.1\pm0.19$	&	$3\nu_5-\nu_6$	\\
$77$	&	$109.331\pm0.014$	&	$0.045\pm0.009$	&	$0.04\pm0.17$	&	$-\nu_2+\nu_4+\nu_6+2\nu_\star$	\\
$78$	&	$100.192\pm0.014$	&	$0.045\pm0.008$	&	$1.06\pm0.16$	&	$\nu_4+\nu_5-\nu_6$	\\
$79$	&	$332.09\pm0.02$	&	$0.04\pm0.01$	&	$1.31\pm0.13$	&	$\nu_1+2\nu_4-\nu_6-2\nu_\star$	\\
$80$	&	$293.430\pm0.016$	&	$0.042\pm0.008$	&	$2.97\pm0.16$	\\
$81$	&	$337.362\pm0.019$	&	$0.044\pm0.008$	&	$6.21\pm0.17$	\\
$82$	&	$12.009\pm0.014$	&	$0.063\pm0.010$	&	$4.21\pm0.11$	\\
$83$	&	$28.812\pm0.012$	&	$0.046\pm0.006$	&	$4.91\pm0.11$	&	$-\nu_3+\nu_4+\nu_6$	\\
$84$	&	$42.061\pm0.017$	&	$0.043\pm0.013$	&	$4.4\pm0.3$	\\
$85$	&	$254.633\pm0.017$	&	$0.046\pm0.007$	&	$0.67\pm0.17$	&	$\nu_3+\nu_4+\nu_5-2\nu_6+2\nu_\star$	\\
$86$	&	$195.270\pm0.013$	&	$0.046\pm0.01$	&	$5.5\pm0.2$	\\
$87$	&	$264.447\pm0.017$	&	$0.042\pm0.008$	&	$4.52\pm0.19$	&	$\nu_1+\nu_3-2\nu_5-\nu_\star$	\\
$88$	&	$129.350\pm0.010$	&	$0.051\pm0.010$	&	$3.41\pm0.17$	&	$\nu_1+\nu_3-3\nu_5-2\nu_\star$	\\
$89$	&	$251.282\pm0.014$	&	$0.043\pm0.009$	&	$2.66\pm0.19$	&	$\nu_1+\nu_3+\nu_4-2\nu_5-\nu_6-\nu_\star$	\\
$90$	&	$165.52\pm0.03$	&	$0.022\pm0.005$	&	$1.7\pm0.4$	&	$\nu_2+\nu_5-\nu_6-\nu_\star$	\\
\hline
\end{tabular}

\end{table}

\begin{table} 
\begin{tabular}{lrrrl}
\hline
\hline
{\bf \#} & {\bf Freq $(\mathrm{\mu Hz})$} & {\bf Amp$(10^{-3})$} & {\bf Phase} \\ \hline
$91$	&	$165.61\pm0.03$	&	$0.025\pm0.007$	&	$0.9\pm0.4$	&	$\nu_2+\nu_5-\nu_6-\nu_\star$	\\
$92$	&	$111.025\pm0.014$	&	$0.043\pm0.008$	&	$4.98\pm0.2$	&	$\nu_3-\nu_6$	\\
$93$	&	$91.454\pm0.017$	&	$0.042\pm0.008$	&	$1.2\pm0.2$	&	$-\nu_2+3\nu_5-\nu_\star$	\\
$94$	&	$49.00\pm0.02$	&	$0.043\pm0.006$	&	$2.02\pm0.14$	&	$-\nu_2+2\nu_4-\nu_5+\nu_6-2\nu_\star$	\\
$95$	&	$271.836\pm0.017$	&	$0.04\pm0.007$	&	$5.5\pm0.3$	&	$\nu_2+\nu_3-\nu_4+2\nu_5-2\nu_6$	\\
$96$	&	$156.11\pm0.02$	&	$0.039\pm0.007$	&	$4.96\pm0.2$	&	$\nu_2+2\nu_3-\nu_4-3\nu_6$	\\
$97$	&	$183.28\pm0.02$	&	$0.04\pm0.01$	&	$5.6\pm0.2$	&	$2\nu_5-2\nu_\star$	\\
$98$	&	$203.639\pm0.02$	&	$0.041\pm0.008$	&	$5.28\pm0.17$	&	$\nu_3+\nu_4-\nu_5-4\nu_\star$	\\
$99$	&	$72.21\pm0.03$	&	$0.036\pm0.009$	&	$6.2\pm0.2$	&	$2\nu_1-\nu_3-\nu_4-\nu_\star$	\\
$100$	&	$49.749\pm0.013$	&	$0.041\pm0.006$	&	$4.2\pm0.2$	&	$\nu_2-\nu_3+4\nu_\star$	\\
$101$	&	$327.468\pm0.019$	&	$0.039\pm0.008$	&	$3.7\pm0.2$	&	$-\nu_2+2\nu_5+2\nu_6+\nu_\star$	\\
$102$	&	$326.98\pm0.03$	&	$0.038\pm0.008$	&	$4.1\pm0.2$	&	$\nu_4+3\nu_5-\nu_6$	\\
$103$	&	$93.402\pm0.019$	&	$0.042\pm0.009$	&	$3.3\pm0.2$	&	$\nu_1+\nu_3-3\nu_4$	\\
$104$	&	$93.21\pm0.02$	&	$0.026\pm0.008$	&	$3.1\pm0.4$	&	$\nu_2-\nu_3-\nu_4+2\nu_5+2\nu_\star$	\\
$105$	&	$92.68\pm0.02$	&	$0.041\pm0.010$	&	$3.76\pm0.16$	&	$\nu_4-\nu_5+\nu_6-4\nu_\star$	\\
$106$	&	$290.41\pm0.05$	&	$0.043\pm0.008$	&	$3.27\pm0.14$	&	$\nu_3+\nu_4-\nu_5$	\\
$107$	&	$314.423\pm0.016$	&	$0.041\pm0.006$	&	$3.1\pm0.2$	&	$2\nu_1-2\nu_5+2\nu_\star$	\\
$108$	&	$248.670\pm0.014$	&	$0.03\pm0.01$	&	$5\pm0.3$	&	$\nu_1-\nu_2+2\nu_5$	\\
$109$	&	$311.435\pm0.019$	&	$0.043\pm0.008$	&	$1.15\pm0.16$	&	$\nu_2+\nu_3-\nu_4+\nu_5-\nu_6$	\\
$110$	&	$165.925\pm0.013$	&	$0.042\pm0.009$	&	$5\pm0.2$	&	$-\nu_2+2\nu_6+4\nu_\star$	\\
$111$	&	$253.663\pm0.017$	&	$0.038\pm0.008$	&	$0.6\pm0.2$	&	$2\nu_2-\nu_5-4\nu_\star$	\\
$112$	&	$294.706\pm0.016$	&	$0.041\pm0.009$	&	$3.4\pm0.2$	&	$-2\nu_2+3\nu_3-2\nu_\star$	\\
$113$	&	$163.78\pm0.02$	&	$0.039\pm0.009$	&	$0.7\pm0.2$	&	$2\nu_1-\nu_3-\nu_5+2\nu_\star$	\\
$114$	&	$120.908\pm0.02$	&	$0.041\pm0.008$	&	$4.13\pm0.19$	&	$\nu_1-\nu_3+\nu_4-\nu_5+\nu_6-2\nu_\star$	\\
$115$	&	$236.78\pm0.02$	&	$0.034\pm0.008$	&	$4.2\pm0.2$	&	$-\nu_2+\nu_4-\nu_5+3\nu_6-\nu_\star$	\\
$116$	&	$170.47\pm0.03$	&	$0.038\pm0.009$	&	$1.6\pm0.2$	&	$-\nu_1+\nu_5+2\nu_6$	\\
$117$	&	$239.813\pm0.016$	&	$0.036\pm0.008$	&	$2.9\pm0.3$	&	$\nu_6+4\nu_\star$	\\
$118$	&	$272.67\pm0.03$	&	$0.033\pm0.008$	&	$2\pm0.2$	&	$-\nu_1-\nu_2+3\nu_3-2\nu_\star$	\\
$119$	&	$313.254\pm0.017$	&	$0.04\pm0.008$	&	$2.7\pm0.2$	\\
$120$	&	$401.797\pm0.017$	&	$0.039\pm0.008$	&	$0.29\pm0.19$	&	$\nu_1+\nu_6$	\\
$121$	&	$42.553\pm0.018$	&	$0.04\pm0.009$	&	$2.4\pm0.2$	&	$2\nu_3-\nu_4+\nu_5-3\nu_6$	\\
$122$	&	$25.576\pm0.013$	&	$0.039\pm0.008$	&	$5.21\pm0.19$	&	$-\nu_4+3\nu_5-\nu_6-\nu_\star$	\\
$123$	&	$28.24\pm0.02$	&	$0.036\pm0.006$	&	$5.58\pm0.19$	&	$-\nu_1+\nu_3-\nu_4+\nu_6$	\\
$124$	&	$24.32\pm0.03$	&	$0.036\pm0.012$	&	$1.5\pm0.2$	&	$\nu_1-\nu_2-\nu_3+\nu_5+\nu_6$	\\
$125$	&	$238.96\pm0.04$	&	$0.033\pm0.009$	&	$3.3\pm0.2$	\\
$126$	&	$47.66\pm0.02$	&	$0.03\pm0.01$	&	$3.51\pm0.2$	&	$\nu_2+2\nu_4-3\nu_6$	\\
$127$	&	$68.073\pm0.014$	&	$0.039\pm0.008$	&	$3.4\pm0.3$	&	$-\nu_1+\nu_3+\nu_4-4\nu_\star$	\\
$128$	&	$289.403\pm0.019$	&	$0.043\pm0.009$	&	$4.09\pm0.17$	&	$\nu_3+\nu_5-\nu_6+3\nu_\star$	\\
$129$	&	$270.149\pm0.018$	&	$0.032\pm0.006$	&	$6.2\pm0.3$	&	$2\nu_5+2\nu_\star$	\\
$130$	&	$332.95\pm0.02$	&	$0.035\pm0.008$	&	$0.3\pm0.2$	&	$3\nu_3-3\nu_6$	\\
$131$	&	$202.43\pm0.02$	&	$0.033\pm0.008$	&	$4.1\pm0.2$	&	$\nu_3-2\nu_4+\nu_6+3\nu_\star$	\\
$132$	&	$31.673\pm0.019$	&	$0.033\pm0.008$	&	$1.1\pm0.3$	\\
$133$	&	$185.22\pm0.02$	&	$0.033\pm0.007$	&	$3.3\pm0.2$	&	$\nu_1+\nu_3-2\nu_6-\nu_\star$	\\
$134$	&	$198.58\pm0.03$	&	$0.034\pm0.009$	&	$3.33\pm0.2$	&	$-\nu_2+\nu_3+\nu_4+\nu_\star$	\\
$135$	&	$286.26\pm0.03$	&	$0.031\pm0.009$	&	$1.51\pm0.2$	&	$\nu_1-\nu_3+2\nu_4+\nu_\star$	\\
\hline
\end{tabular}
\end{table}

\begin{table}
\begin{tabular}{lrrrl}
\hline
\hline
{\bf \#} & {\bf Freq $(\mathrm{\mu Hz})$} & {\bf Amp$(10^{-3})$} & {\bf Phase} \\ \hline
$136$	&	$137.145\pm0.017$	&	$0.031\pm0.009$	&	$2.7\pm0.3$	&	$-\nu_2+\nu_3-2\nu_4+2\nu_5+\nu_6$	\\
$137$	&	$182.74\pm0.02$	&	$0.035\pm0.008$	&	$1.6\pm0.3$	&	$\nu_1-\nu_6+4\nu_\star$	\\
$138$	&	$312.192\pm0.02$	&	$0.031\pm0.007$	&	$1.6\pm0.3$	&	$\nu_3+\nu_5-3\nu_\star$	\\
$139$	&	$217.84\pm0.02$	&	$0.029\pm0.007$	&	$6\pm0.3$	&	$\nu_1+\nu_2-2\nu_4+\nu_\star$	\\
$140$	&	$84.76\pm0.02$	&	$0.032\pm0.008$	&	$2\pm0.2$	&	$\nu_3+2\nu_4-3\nu_6$	\\
$141$	&	$395.06\pm0.02$	&	$0.032\pm0.008$	&	$5.3\pm0.2$	&	$\nu_1+\nu_2-\nu_3+2\nu_5-2\nu_\star$	\\
$142$	&	$148.59\pm0.03$	&	$0.033\pm0.009$	&	$2.5\pm0.3$	&	$-2\nu_1+\nu_2+3\nu_4$	\\
$143$	&	$193.90\pm0.03$	&	$0.031\pm0.008$	&	$5\pm0.3$	&	$-\nu_2+\nu_3+\nu_5+2\nu_\star$	\\
$144$	&	$192.95\pm0.02$	&	$0.032\pm0.008$	&	$5.5\pm0.2$	\\
$145$	&	$166.96\pm0.03$	&	$0.034\pm0.009$	&	$3.18\pm0.2$	&	$-\nu_2+3\nu_6-3\nu_\star$	\\
$146$	&	$38.34\pm0.02$	&	$0.035\pm0.009$	&	$3.5\pm0.2$	&	$2\nu_2-3\nu_6+2\nu_\star$	\\
$147$	&	$197.866\pm0.014$	&	$0.034\pm0.009$	&	$5.3\pm0.3$	&	$\nu_3-\nu_6+4\nu_\star$	\\
$148$	&	$67.21\pm0.03$	&	$0.031\pm0.008$	&	$3.58\pm0.19$	\\
$149$	&	$364.973\pm0.019$	&	$0.031\pm0.008$	&	$3\pm0.3$	&	$\nu_1-2\nu_2+\nu_3+2\nu_6$	\\
$150$	&	$287.403\pm0.013$	&	$0.043\pm0.008$	&	$1.21\pm0.17$	&	$2\nu_1+\nu_2-3\nu_6+\nu_\star$	\\
$151$	&	$79.76\pm0.02$	&	$0.029\pm0.007$	&	$1.9\pm0.3$	&	$\nu_3-2\nu_4+2\nu_5-\nu_6+\nu_\star$	\\
$152$	&	$211.260\pm0.014$	&	$0.032\pm0.008$	&	$5.9\pm0.3$	&	$-\nu_1+3\nu_2-\nu_3+2\nu_\star$	\\
$153$	&	$72.78\pm0.02$	&	$0.029\pm0.007$	&	$4.2\pm0.3$	&	$\nu_1-\nu_3+\nu_6-3\nu_\star$	\\
$154$	&	$273.223\pm0.019$	&	$0.03\pm0.009$	&	$0.5\pm0.4$	&	$2\nu_1-\nu_2-\nu_3+\nu_5+\nu_6$	\\
$155$	&	$86.73\pm0.02$	&	$0.03\pm0.008$	&	$4.1\pm0.3$	&	$4\nu_\star$	\\
$156$	&	$228.637\pm0.012$	&	$0.034\pm0.008$	&	$5.9\pm0.3$	\\
$157$	&	$335.35\pm0.02$	&	$0.03\pm0.008$	&	$1\pm0.3$	&	$\nu_2+2\nu_3-3\nu_4$	\\
$158$	&	$204.75\pm0.03$	&	$0.031\pm0.008$	&	$4.7\pm0.3$	&	$-\nu_1+\nu_2+2\nu_5$	\\
$159$	&	$61.35\pm0.02$	&	$0.032\pm0.008$	&	$5.2\pm0.3$	&	$-\nu_5+\nu_6+\nu_\star$	\\
$160$	&	$298.78\pm0.03$	&	$0.03\pm0.008$	&	$3.3\pm0.3$	&	$2\nu_1-\nu_3+3\nu_\star$	\\
$161$	&	$287.908\pm0.014$	&	$0.04\pm0.012$	&	$0.1\pm0.3$	\\
$162$	&	$292.95\pm0.02$	&	$0.03\pm0.009$	&	$5.1\pm0.3$	&	$3\nu_1-\nu_2-2\nu_5$	\\
$163$	&	$284.25\pm0.02$	&	$0.029\pm0.008$	&	$2.4\pm0.2$	&	$2\nu_6-\nu_\star$	\\
$164$	&	$249.110\pm0.018$	&	$0.032\pm0.008$	&	$3.8\pm0.3$	&	$2\nu_1-\nu_2-\nu_\star$	\\
$165$	&	$51.09\pm0.03$	&	$0.03\pm0.008$	&	$2.3\pm0.3$	&	$\nu_1-\nu_3+\nu_6-4\nu_\star$	\\
$166$	&	$52.63\pm0.02$	&	$0.029\pm0.008$	&	$1.8\pm0.3$	&	$-\nu_2+2\nu_3-2\nu_5-\nu_\star$	\\
$167$	&	$22\pm0.03$	&	$0.028\pm0.008$	&	$2.9\pm0.3$	&	$\nu_1-\nu_2$	\\
$168$	&	$255.43\pm0.04$	&	$0.026\pm0.008$	&	$2.5\pm0.3$	&	$2\nu_1-\nu_3+\nu_\star$	\\
$169$	&	$88.646\pm0.02$	&	$0.031\pm0.008$	&	$2.2\pm0.3$	\\
$170$	&	$153.45\pm0.03$	&	$0.029\pm0.009$	&	$2.2\pm0.3$	&	$\nu_1-2\nu_5+\nu_6-\nu_\star$	\\
$171$	&	$401.42\pm0.03$	&	$0.025\pm0.007$	&	$3.2\pm0.4$	&	$2\nu_5+\nu_6+\nu_\star$	\\
$172$	&	$280.06\pm0.04$	&	$0.025\pm0.008$	&	$4.4\pm0.4$	&	$\nu_1+2\nu_4-2\nu_5-\nu_\star$	\\
$173$	&	$188.12\pm0.02$	&	$0.028\pm0.008$	&	$6\pm0.3$	&	$\nu_4+\nu_5-3\nu_\star$	\\
$174$	&	$266.31\pm0.03$	&	$0.026\pm0.009$	&	$0\pm0.3$	&	$\nu_5+\nu_6$	\\
$175$	&	$301.43\pm0.03$	&	$0.026\pm0.008$	&	$3.1\pm0.3$	&	$\nu_4+2\nu_5-3\nu_\star$	\\
$176$	&	$10.194\pm0.019$	&	$0.04\pm0.006$	&	$2.29\pm0.16$	\\
$177$	&	$13.483\pm0.014$	&	$0.043\pm0.007$	&	$5.8\pm0.2$	&	$3\nu_1-\nu_2-2\nu_3+\nu_\star$	\\
$178$	&	$30.285\pm0.018$	&	$0.035\pm0.009$	&	$0.2\pm0.3$	&	$\nu_4-\nu_6+2\nu_\star$	\\
$179$	&	$21.4\pm0.02$	&	$0.038\pm0.007$	&	$2\pm0.17$	&	$-\nu_1+\nu_2+2\nu_\star$	\\
$180$	&	$17.33\pm0.02$	&	$0.036\pm0.010$	&	$1.18\pm0.17$	\\
\hline
\end{tabular}
\end{table}

\begin{table}
\begin{tabular}{lrrrl}
\hline
\hline
{\bf \#} & {\bf Freq $(\mathrm{\mu Hz})$} & {\bf Amp$(10^{-3})$} & {\bf Phase} \\ \hline
$181$	&	$17.74\pm0.03$	&	$0.033\pm0.009$	&	$5.3\pm0.2$	&	$-3\nu_4+3\nu_6-\nu_\star$	\\
$182$	&	$27.15\pm0.02$	&	$0.038\pm0.009$	&	$3.2\pm0.2$	&	$\nu_1-3\nu_2+3\nu_6$	\\
$183$	&	$16.10\pm0.04$	&	$0.025\pm0.007$	&	$4\pm0.3$	&	$\nu_1-\nu_3+2\nu_4-2\nu_5-\nu_\star$	\\
$184$	&	$258.35\pm0.02$	&	$0.03\pm0.007$	&	$3.8\pm0.3$	\\
$185$	&	$1.153\pm0.006$	&	$0.31\pm0.03$	&	$5.00\pm0.09$	\\
$186$	&	$9.266\pm0.008$	&	$0.119\pm0.016$	&	$2.81\pm0.08$	&	$-3\nu_5+2\nu_6+2\nu_\star$	\\
$187$	&	$1.880\pm0.003$	&	$0.20\pm0.02$	&	$3.70\pm0.03$	\\
$188$	&	$0.23\pm0.14$	&	$0.197\pm0.008$	&	$3.923\pm0.003$	\\
$189$	&	$3.621\pm0.008$	&	$0.171\pm0.016$	&	$2.49\pm0.08$	&	$-\nu_2+3\nu_5-\nu_6+2\nu_\star$	\\
$190$	&	$2.188\pm0.003$	&	$0.144\pm0.019$	&	$4.75\pm0.06$	&	$-\nu_2-\nu_3+3\nu_5+\nu_6$	\\
$191$	&	$3.898\pm0.012$	&	$0.062\pm0.014$	&	$0.74\pm0.10$	&	$-\nu_4+3\nu_5-\nu_6-2\nu_\star$	\\
$192$	&	$211.962\pm0.014$	&	$0.044\pm0.007$	&	$4.2\pm0.17$	&	$\nu_3-\nu_4+\nu_6-3\nu_\star$	\\
$193$	&	$6.81\pm0.014$	&	$0.043\pm0.007$	&	$1.6\pm0.2$	&	$-\nu_2+\nu_3-\nu_4+\nu_6-2\nu_\star$	\\
$194$	&	$2.97\pm0.014$	&	$0.056\pm0.008$	&	$4.6\pm0.07$	&	$-\nu_2+\nu_3-3\nu_5+2\nu_6$	\\
$195$	&	$288.790\pm0.018$	&	$0.04\pm0.008$	&	$3.2\pm0.2$	&	$-\nu_2+\nu_4+\nu_5+2\nu_6-2\nu_\star$	\\
$196$	&	$1.205\pm0.006$	&	$0.04\pm0.006$	&	$0.75\pm0.05$	\\
$197$	&	$130.063\pm0.019$	&	$0.036\pm0.008$	&	$2.4\pm0.2$	&	$\nu_1+3\nu_5-3\nu_6$	\\
$198$	&	$268.76\pm0.02$	&	$0.036\pm0.008$	&	$4.6\pm0.2$	&	$\nu_3+\nu_4-\nu_5-\nu_\star$	\\
$199$	&	$7.19\pm0.03$	&	$0.036\pm0.007$	&	$4.69\pm0.17$	&	$3\nu_1-2\nu_2-\nu_3-\nu_\star$	\\
$200$	&	$174.33\pm0.02$	&	$0.032\pm0.008$	&	$2.4\pm0.3$	&	$3\nu_1-\nu_5-3\nu_6$	\\
$201$	&	$23.8\pm0.02$	&	$0.033\pm0.008$	&	$3.7\pm0.2$	\\
$202$	&	$137.53\pm0.02$	&	$0.032\pm0.009$	&	$1.2\pm0.3$	&	$\nu_3+\nu_5-\nu_6-4\nu_\star$	\\
$203$	&	$12.81\pm0.02$	&	$0.032\pm0.007$	&	$2.6\pm0.2$	&	$2\nu_2-\nu_5-2\nu_6-\nu_\star$	\\
$204$	&	$8.522\pm0.017$	&	$0.032\pm0.007$	&	$5.7\pm0.3$	&	$-2\nu_1+2\nu_3-\nu_\star$	\\
\hline
\end{tabular}
\caption{The best-fit frequencies, amplitudes, and phases of 204 statistically significant oscillations in KOI-976's short-cadence \emph{Kepler} photometry. We subtract each oscillation using the Linear Algorithm for Significance Reduction ({\tt LASR}) and calculate uncertainties following \citet{ahlers2018lasr}. We also identify possible combination frequencies of the six highest-amplitude frequencies and the rotation frequency. We find that 160 out of 204 total oscillations match combination values of KOI-976's dominant modes of oscillation and identified rotation frequency. All predicted combinations are within $1\sigma$ of their observed values, with the average uncertainty of all combinations below $0.5\sigma$. }
\label{table:oscillations}
\end{table}

\end{document}